\documentclass[fleqn,usenatbib]{mnras}

\usepackage[T1]{fontenc}
\DeclareRobustCommand{\VAN}[3]{#2}
\let\VANthebibliography\thebibliography
\def\thebibliography{\DeclareRobustCommand{\VAN}[3]{##3}\VANthebibliography}

\usepackage{graphicx}	
\usepackage{amsmath}	
\usepackage{amssymb}	
\usepackage[flushleft]{threeparttable}
\usepackage[stable]{footmisc}

\newcommand{\mdot}{\ensuremath{\dot{m}/\dot{m}_{\rm Edd}}}
\newcommand{\rg}{\ensuremath{r_{\rm g}}}
\newcommand{\ltransf}{\ensuremath{L_{\rm transf}/L_{\rm disc}}}
\newcommand{\fcol}{\ensuremath{f_{\rm col}}}
\newcommand{\Lxfit}{\ensuremath{L_{\rm 2-10, fit}/L_{\rm Edd}}}
\newcommand{\Lxobs}{\ensuremath{L_{\rm 2-10, obs}/L_{\rm Edd}}}

\title[Revisiting UV/optical continuum time lags in AGN]{Revisiting UV/optical continuum time lags in AGN}

\author[E. S. Kammoun et al.]{
 E. S. Kammoun,$^{1,2}$\thanks{E-mail: \href{mailto:ekammoun@irap.omp.eu}{ekammoun@irap.omp.eu}} 
 L. Robin,$^{3,1}$
I. E. Papadakis,$^{4,5}$
M. Dov{\v c}iak,$^{6}$
C. Panagiotou$^{7}$
\\
$^{1}$IRAP, Universit\'{e} de Toulouse, CNRS, UPS, CNES, 9, Avenue du Colonel Roche, BP 44346, F-31028, Toulouse Cedex 4, France \\
$^{2}$INAF -- Osservatorio Astrofisico di Arcetri, Largo Enrico Fermi 5, I-50125 Firenze, Italy\\
$^{3}$Institut Sup\'{e}rieur de l’A\'{e}ronautique et de l’Espace, 31400 Toulouse, France\\
$^{4}$Department of Physics and Institute of Theoretical and Computational Physics, University of Crete, 71003 Heraklion, Greece\\
$^{5}$Institute of Astrophysics, FORTH, GR-71110 Heraklion, Greece\\
$^{6}$Astronomical Institute of the Czech Academy of Sciences, Bo{\v c}n{\'i} II 1401, CZ-14100 Prague, Czech Republic\\
$^{7}$MIT Kavli Institute for Astrophysics and Space 2 Research, Massachusetts Institute of Technology, Cambridge, MA 02139, USA
}

\date{Accepted XXX. Received YYY; in original form ZZZ}

\pubyear{2022}

\begin{document}
\label{firstpage}
\pagerange{\pageref{firstpage}--\pageref{lastpage}}
\maketitle

\begin{abstract}

In this paper, we present an updated version of our model ({\tt KYNXiltr}) which considers thermal reverberation of a standard Novikov-Thorne accretion disc illuminated by an X–ray point-like source. Previously, the model considered only two cases of black hole spins, and assumed a colour correction factor $\fcol = 2.4$. Now, we extend the model to any spin value and \fcol. In addition, we consider two scenarios of powering the X--ray corona, either via accretion, or external to the accretion disc. We use {\tt KYNXiltr} to fit the observed time lags obtained from intense monitoring of four local Seyfert galaxies (NGC\,5548, NGC\,4593, Mrk\,817, and Fairall\,9). We consider various combinations of black hole spin, colour correction, corona height, and fraction of accretion power transferred to the corona. The model fits well the overall time-lags spectrum in these sources (for a large parameter space). For NGC\,4593 only, we detect a significant excess of delays in the U-band. The contribution of the diffuse BLR emission in the time-lags spectrum of this source is significant. It is possible to reduce the large best-fitting parameter space by combining the results with additional information, such as the observed Eddington ratio and average X–ray luminosity. We also provide an update to the analytic expression provided by Kammoun et al., for an X--ray source that is not powered by the accretion process, which can be used for any value of \fcol, and for two values of the black hole spin (0 and 0.998).

\end{abstract}
\begin{keywords}
accretion, accretion discs -- galaxies: nuclei -- galaxies: Seyferts -- X--rays: individual: NGC\,5548, NGC\,4593, Mrk\,817, Fairall\,9
\end{keywords}



\section{Introduction}
\label{sec:intro}

The current paradigm assumes that Active Galactic Nuclei (AGN) are powered by the accretion of matter onto a supermassive back hole (SMBH) from a geometrically thin and optically thick disc. Thermal UV photons emerge from the disc and are Compton up scattered in a medium of hot electrons, known as the corona, and emitted in X--rays \citep[e.g.,][]{Lightman88, Haardt93}. Assuming an isotropic emission, part of theses photons are directly emitted in the direction of the observer. The other part will illuminate back the accretion disc, where they will be partially reflected in X--rays \citep[e.g.,][]{George91, Matt1991} and partially absorbed and then re-emitted in the UV/optical range in the form of thermalised emission. As the X--ray source varies in time, correlated variability should be detected in the UV/optical band with a time-lag depending on the wavelength \citep[e.g.,][]{Cackett07}. 

Intense, multi-wavelength, monitoring campaign using space-based and ground-based telescopes confirmed the presence of correlation between AGN light curves in various UV/optical bands where the time delay increases as function of wavelength \citep[e.g.,][]{Edelson15, Fausnaugh16, Mchardy18, Cackett18, Edelson19, Cackett20, Santisteban20, Pahari20, Kara21, Vincentelli21, Vincentelli22, McHardy23, Kara23, Donnan23}. In many of these cases modelling the data assuming thermal reverberation was able to explain the shape of the time-lags as function of wavelength (time-lag spectra), while under-predicting its amplitude \citep[see e.g.,][]{Fausnaugh16}. However, the model thermal reverberation time lags were calculated using basic approximations like the viscous heating and the assumption that X--rays contribute equal amounts of energy to the disc at all radii. In addition, these models estimate the time lags by simply considering the light-travel delay radius emitting light at a wavelength $\lambda$, without considering the source height and general relativity effects due to the presence of the central black hole (BH), while the radius is computed by assuming Wien's displacement law. This is a law which can predict the peak wavelength ($\lambda_{\rm peak}$) assuming a certain blackbody temperature, but it cannot give the correct temperature given $\lambda$ (obviously the temperature would be different if instead of $\lambda$ one would consider the frequency $\nu$). In addition, the previous models would not consider BH spins other than zero. Despite these short comings, results from model fits to the observed time-lags were accepted as an indication that the discs are larger than expected, requiring accretion rates that are larger than the ones inferred from multi-wavelength analysis and broadband spectral-energy distribution (SED) fitting.

In recent studies, \citet[][hereafter K21b]{Kammoun19lag, Kammoun21a} presented {\tt KYNXiltr}, a model able to compute the response functions ($\Psi$) of Novikov-Thorne (NT) accretion discs \citep{Novikov73} illuminated by a lamp-post X--ray corona, taking into account all general relativity and disc ionization effects. These responses were used then to estimate the average time delay at a given wavelength ($\lambda)$. K21b studied the effect of various AGN parameters on the time lags, and presented an analytic prescription that can be used to model the observed time-lag spectra. This was then used by \citet[][hereafter K21a]{Kammoun21b} to successfully model the time-lag spectra in seven AGNs. The estimated accretion rate required to fit the time-lag spectra were in good agreement with literature. \citet{Panagiotou20, Panagiotou22} used the same model to compute UV/optical power spectral densities (PSD) and were able to model the PSD from the long monitoring campaign of NGC\,5548. Recently, \citet{Dovciak22} presented a new model ({\tt KYNSED}) of the SED around accreting black holes considering the X--ray irradiation of an NT accretion disc by a lamp-post corona. {\tt KYNSED} considers the case where the X--ray luminosity is independent of the accretion power of the disc, and the case where the X--ray luminosity is equal to the accretion power generated in the disc within a radius, $r_{\rm transf}$. The X--ray luminosity is then parameterised by the ratio of the accretion power within $r_{\rm transf}$ to the total accretion power, \ltransf. 

In this work, we present a new code {\tt KYNXiltr}\footnote{The code is publicly available at \url{https://projects.asu.cas.cz/dovciak/kynxiltr}. We present also a detailed description of how the code can be used to perform simulations of time lags or fit observed time lag spectra.}  which can be used to fit the observed time-lag spectra. There are a few differences between the new code and the approach of K21b. While the analytical time-lags equations presented by K21b assume that the X--ray luminosity is not part of the power that is liberated by the accretion process in the disc. The new code can compute time lags under this assumption but also under the assumption of the X--ray luminosity being equal to the accretion power generated within a certain radius, $r_{\rm transf}$. Moreover, the model time lags in K21b were computed assuming a colour correction factor of $\fcol = 2.4$ for the disc emission and that the disc extends from the innermost stable circular orbit (ISCO) to a (fixed) outer radius of $R_{\rm out} = 10^4\,\rg$. In the new code, time-lags can be computed for any color correction factor and for any disc outer radius. Finally, K21b considered only two spins, namely a spin of 1 and zero, while the new code we present in this work can  compute time-lags for any spin value, from 0 up to 1.

In Section\,\ref{sec:model_setup}, we describe how the new code works, and we present a discussion of the time-lags dependence on the new parameters we introduce in our model. In Section\,\ref{sec:sample}, we introduce the AGN sample that we use to fit their observed time-lags. In Section\,\ref{sec:lagfit} we present the fitting technique we use and the results. Finally, we discuss the results and summarize our work in Section\,\ref{sec:conclusions}.

\section{Modeling the X--ray thermal reverberation of the disc}
\label{sec:model_setup}
\begin{figure*}
    {\centering
    \includegraphics[width=0.95\textwidth]{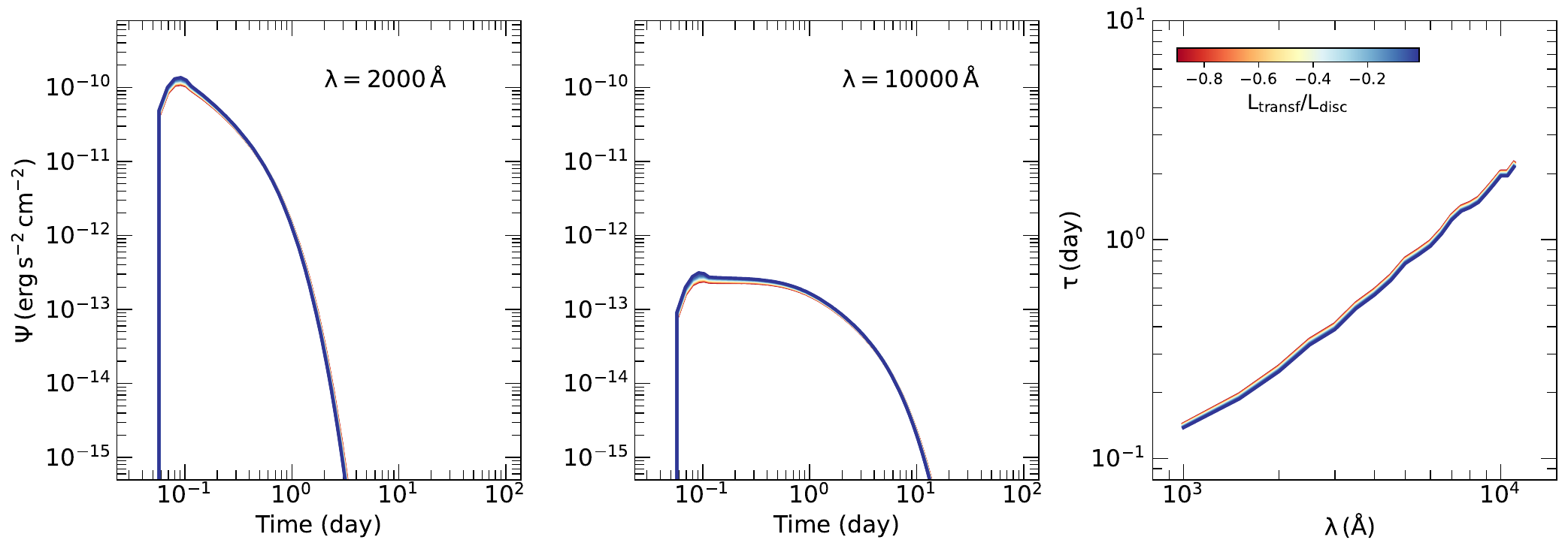}
    \includegraphics[width=0.95\textwidth]{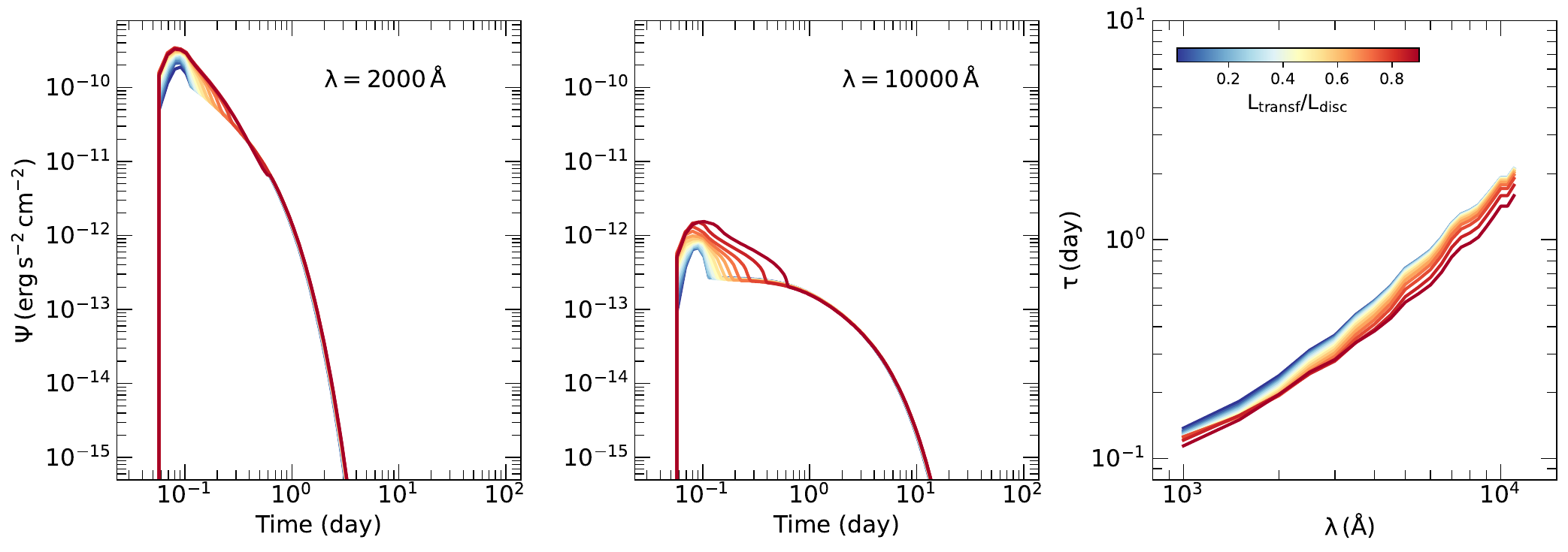} 
    \includegraphics[width=0.95\textwidth]{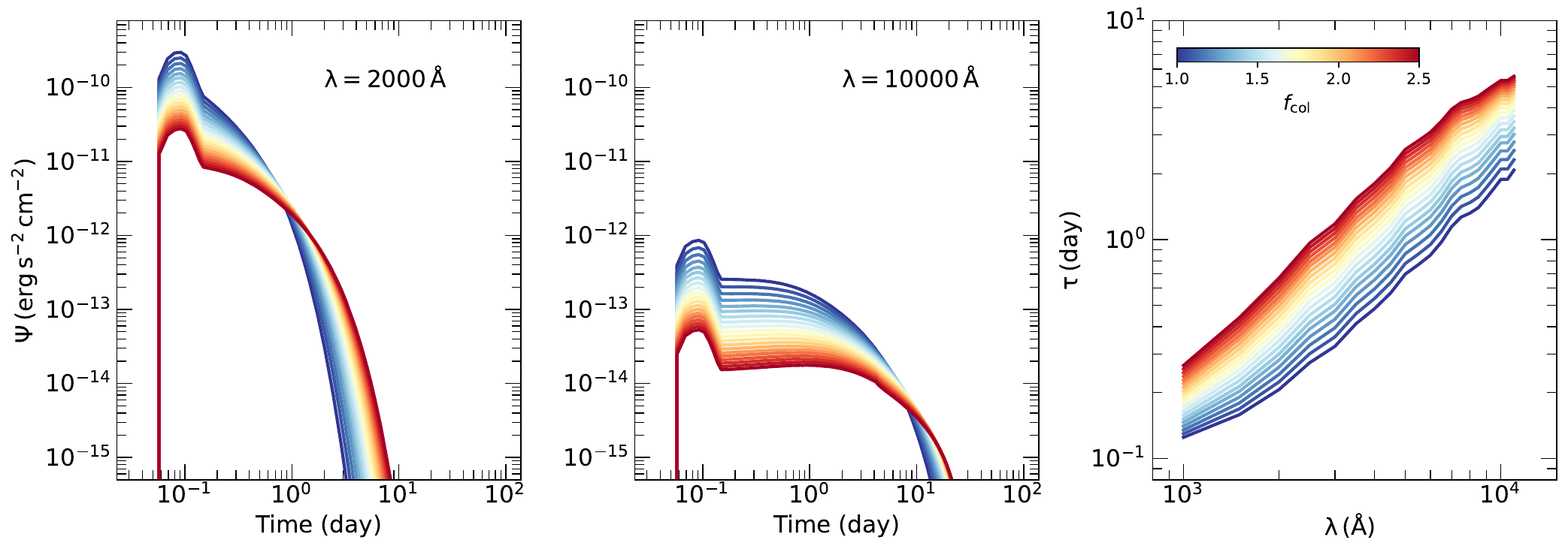}
    }
\caption{Response functions at 2000\,\AA\ and 10000\,\AA\ (left and middle panels, respectively) and time lag spectra (right panels) for different negative values \ltransf\ (top), positive values of \ltransf\ (center), and \fcol\ (bottom). The fiducial parameters are assumed to be $a^\ast = 0.7$, $M_{\rm BH} = 5\times10^7\,\rm M_\odot$, $h = 10\, \rm r_g$, $\dot{m}_{\rm Edd} = 0.05$, $\Gamma = 2$, $E_{\rm cut} = 300\, \rm keV$, $R_{\rm out} = 5000 \, \rm r_g$. In the case of different values of \ltransf, we fixed \fcol\ at 1. In the case of different values of \fcol\, we fixed \ltransf\ at 0.5.}
\label{fig:response_lag}
\end{figure*}


Similar to K21b, we  assume a lamp-post X--ray corona at a given height ($h$) on the rotational axis of the black hole, which emits isotropically in its rest frame a power-law spectrum of the form $f_X(t)=N(t)E^{-\Gamma} \rm{exp}\left(-E/E_{cut}\right)$,  where $\Gamma$ is assumed to be constant. We need to determine the disc response in order to model the the time-lags when the disc is illuminated by variable X-rays. We provide below a short description of how this is done \citep[a detailed description is given by] []{Kammoun21a}.

Part of the X--ray flux received by the disc will be reflected and re-emitted in X-rays (this is the “disc reflection component”), and part of it will be absorbed. The absorbed X-rays will thermalise in the disc. They will act as an extra source of heating hence the local disc temperature, and the disc emission, will increase. As a result, the total disc flux, as observed by a distant observer, will vary with time because the X-ray flash first illuminates the inner disc and then propagates to the outer parts. 

The disc response function, $\Psi(\lambda,t_{\rm obs})$,  is equal to the flux that the disc emits due to X-ray heating at wavelength $\lambda$, and at time $t_{\rm obs}$, as measured by a distant observer. We identify all disc elements that brighten up at time  $t_{\rm obs}$, for the observer, we compute the sum of their flux (say $F_{\rm tot}(\lambda, t_{\rm obs})$), we subtract the sum of their intrinsic disc flux, say $F_{\rm NT, t_{obs}}(\lambda)$ \citep[NT stands for the disc flux  calculated following][]{Novikov73}, and then the disc response is defined as, 

\begin{equation}
    \Psi(\lambda,t_{\rm obs})\propto \frac{F_{\rm tot}(\lambda, t_{\rm obs})-F_{\rm NT,t_{obs}}(\lambda)}{L_{\rm X}}.
\label{eq:psidefine}
\end{equation}

\noindent Once the disc response is known, then the disc flux that a distant observer will detect in the UV/optical bands, when the disc is constantly being illuminated by variable X-rays, will be given by,

\begin{equation}
    F_{obs}(\lambda, t)= F_{\rm NT}(\lambda) +\int_0^{\infty} L_{\rm X}(t-t')\Psi_{L_{\rm X}}(\lambda,t') dt',
\end{equation}

\noindent where $F_{\rm obs}(\lambda, t)$ is the flux at wavelength $\lambda$ emitted by the whole disc at time $t$, and $F_{\rm NT}(\lambda t)$, is the flux emitted at $\lambda$ by an NT disc.

The response function determines, to a large extent, the cross-correlation between the X--ray and the UV/optical variations. In fact, the centroid of the response function, should be representative of the maximum peak in the cross-correlation function (CCF) that is measured between the X-ray and the UV/optical light curves. The centroid is defined as follows,
\begin{equation}\label{eq:tau}
    \tau(\lambda) = \frac{\int t \Psi(t,\lambda) {\rm dt}}{ \int \Psi(t,\lambda)\rm dt}.
\end{equation}
\noindent 

\noindent Like K21b, {\tt KYNXiltr} uses the equation above to compute model time-lags. However, there are some differences between K21b and the new code, which we describe below. 

K21b used the observed $2-10$\,keV luminosity of the corona (in Eddington units), $L_{\rm Xobs,Edd}(t)$, to normalize the response function in Eq.\,\ref{eq:psidefine}. In general, $L_{\rm X}$ in this equation can be any quantity representative of the X--ray corona luminosity. In this work, we follow \cite{Dovciak22} and we 
use the total luminosity of the corona to normalize the disc response function. We parameterize $L_{\rm X}$ as the ratio of the total X--ray luminosity over the accretion power, \ltransf. If \ltransf\ is positive, then $L_{\rm X}$ is equal to the power that is released by the accretion process below a radius, which is transferred to the corona by an unknown physical mechanism. The total $L_{\rm X}$ is this case is equal to (\ltransf)$L_{\rm disc}$, where $L_{\rm disc}$ is set by the accretion rate, $\dot{M}$, as: $L_{\rm disc}=\eta \dot{M}c^2$ ($\eta$ being the accretion efficiency). A negative value of \ltransf\ would mean that the power given to the corona is external to the power released by the accretion power in the disc. In this case, \ltransf\ can be larger than unity, contrary to the case when $\ltransf>0$, when it cannot be larger than 1 (in fact, the code does not allow the use of  a value larger than 0.9 for \ltransf.)

The disc fluxes in Eq.\,\ref{eq:psidefine} are modelled by a color-corrected black body of the form 

\begin{equation}
    I_\nu = \frac{2h}{c^2f_{col}^4}\frac{\nu^3}{\exp(\frac{h\nu}{f_{col}kT})-1},
\end{equation}

\noindent where $I_\nu$ is the specific intensity, $k$ is
Boltzmann’s constant, $h$ is Planck’s constant, and $T$ is the disc temperature. In the case of $F_{\rm NT,t_{obs}}(\lambda)$
the temperature of the disc elements that an observer detects at $t_{\rm obs}$ is set according to NT. In the case of $F_{\rm tot}(\lambda,t_{obs})$ we use the new disc temperature, which is computed after adding to the local disc flux the X--ray flux absorbed by the disc. For both terms though, the emitted spectrum is not equal to that of a blackbody.  It is well known that electron scattering plays
an important role and can lead to deviations from blackbody emission. In general, transfer of energy between electrons and photons that are Compton scattered enforces a Wien tail at the high energy end of the spectrum. The resulting spectra can be approximately modelled by the equation above, where \fcol\ is the multiplicative factor by which spectral features are shifted to higher energies, while \fcol$^{-4}$ keeps the frequency integrated flux fixed. K21b assumed that $\fcol=2.4$, following \cite{Ross92}. The new code can provide time-lags for any \fcol\ value, including the prescription of \cite{Done2012}, when $\fcol=-1$.

An additional improvement of {\tt KYNXiltr} when compared to the analytic expressions of K21b is related with the outer disc radius, $R_{\rm out}$. Although K21b had studied in detail the effects of the disc outer radius to the time-lags (see their Section\,3.5), the analytic expressions presented by K21b were computed assuming an outer radius fixed at $10^4\,\rg$. The new code can compute time-lags from any outer disc radius, as this  value may affect significantly the observed time-lags \citep[see for example][]{McHardy23}. Finally, K21b presented analytic expressions for the time-lags only in the case of a non rotating and a maximally rotating black hole, while {\tt KYNXiltr} can compute time lags for any BH spin value.

We investigated the effects of the BH mass, spin, accretion rate, corona height, disc inclination, and photon index on the response functions and the time lags, using {\tt KYNXiltr}. The effects of these parameters are identical to the ones described in detail by K21b. In the following, we present only the effects of \ltransf\ and \fcol\ on thermal reverberation time lags, as these are the only two parameters that were not studied by K21b. 

\subsection{Dependence of the time lags on \ltransf}
\label{sec:Ltransfdep}

Figure\,\ref{fig:response_lag} shows the effect of changing \ltransf\ on the response functions at 2000\,\AA\ and 10000\,\AA, and the time-lag spectra for negative and positive values of \ltransf\ (upper and middle panels, respectively), and for different values of \fcol\ (bottom panels). These simulations are performed for  $a^\ast = 0.7$, $M_{\rm BH} = 5\times10^7\,\rm M_\odot$, $h = 10\, \rm r_g$, $\dot{m}_{\rm Edd} = 0.05$, $\Gamma = 2$, $E_{\rm cut} = 300\, \rm keV$, $R_{\rm out} = 5000 \, \rm r_g$. In the case of different values of \ltransf, we fixed \fcol\ at 1. In the case of different values of \fcol\, we fixed \ltransf\ at 0.5.

In the case of negative \ltransf\, (which is the case in K21b) the effect is similar to changing the X--ray luminosity in K21b. Our results confirm the non-linearity of the disc response. This effect is clearer in Fig.\,15 of K21b, as for this case, we consider a wider range of X--ray luminosity to better highlight this effect. The observed $2-10\,\rm keV$ luminosity considered in K21b range between 0.001 and 0.5 times the Eddington luminosity. However, in this work, given the considered values \ltransf, this translates into observed $2-10\,\rm keV$ luminosity between 0.0004 and 0.003 times the Eddington luminosity. We recall that the disc response is normalised to the observed X--ray luminosity (see Eq.\,3 in K21b). In this case, if the response of the disc scales linearly with the X--ray luminosity, we would expect the response functions for the various X--ray luminosity to overlap (in all bands), which is not the case. The response functions (at all wavelengths) decrease in amplitude and broaden in time as \ltransf\ increases. This is due to the fact that the thermalised flux does not contribute to the response in each waveband in the same way at all times. In addition, the increase in luminosity leads to an increase in the ionization state of the disc, especially in the inner regions, which will affect the reverberated flux. As a result, the increase in \ltransf (for the negative values) leads to an increase in the time lags at all wavelengths. These effects are detailed in Section\,3.6 of K21b.

In the case of positive values of \ltransf, i.e., when the X--rays are powered by extracting energy from the inner parts of the accretion disc, early times are dominated by the reprocessed emission from illuminating the innermost regions of the disc, within which the power is extracted. Contrary to the case of the externally powered X--ray source, the emission from this part is broader and brighter for larger \ltransf. As \ltransf\ increases the radius $r_{\rm transf}$ within which the accretion power is transferred to the X--ray source increases which leads to a broader disc response from that region. In addition, the X--ray luminosity increases (assuming the same height) which leads to an increase in the amplitude of the reprocessed emission. As the time passes, we detect the emission from the outer parts of the disc that gets fainter with time. The peak due to the illumination of the innermost parts of the disc shifts the centroid of the response function towards smaller times, which leads to a decrease in the time-lag as \ltransf\ increases {\it which, again, is contrary to what happens in the case of the externally illuminated X--ray source}. This effect is amplified at longer wavelengths.

We note that the response function for the case of negative and positive \ltransf\ are identical at times when we observe the thermal reverberation from large radii ($r > r_{\rm transf}$). At smaller times, the amplitude of $\Psi$ is larger in the case when the X--ray source is powered by the accretion process. This difference is due to the fact that the response function is estimated as the difference between the total observed flux and the intrinsic NT disc flux. In the case of accretion powered X--ray source, the intrinsic disc emission is zero, which leads to larger value of $\Psi$ compared to the case where the X--ray corona is not powered by accretion. This difference in $\Psi$ at short time scales shifts the time lags to smaller values in the case where the corona is powered by the accretion process for a given value of \ltransf. The difference in time lags between the two cases increases as the X--ray luminosity increases, to reach $\sim 60\%$. It is also worth noting that the range of time lags for different values of \ltransf\ is larger in the case of a corona powered by the accretion process.

\subsection{Dependence of the time lags on \fcol\ }

The bottom panel of Fig.\,\ref{fig:response_lag} shows the effect of changing \fcol\ on the response functions and the time lag spectra. We fixed all the parameters to the same values as before, considered $\ltransf = 0.5 $, and we varied \fcol\ between 1 and 2.5. The response functions decrease in amplitude and get broader for larger values of \fcol. This effect is similar to an increase in \mdot. In fact, as \fcol\ increases the observed temperature at each disc element increases. In addition, the total observed flux is re-normalised by $\fcol^4$ in order to conserve the total emitted flux. For these reasons, as \fcol\ increases the response function (considered as the difference between the total observed flux and the NT flux at a given wavelength) decreases in amplitude. Thus, the thermalised flux at a given wavelength will be emitted from outer regions of the disc and thus will last longer. This explains the fact that the response functions last longer (i.e. are broader) for larger \fcol.  As a result, the time lags increase as \fcol\ increases.

\section{Sample}
\label{sec:sample}

In order to demonstrate how well the new code works in practice, we chose to fit the time lags of NGC\,4593, NGC\,5548, Fairall\,9, and  Mrk\,817. The time lags for these sources have been determined using long, densely sampled light curves at many wavelengths. Table\,\ref{tab:sources} list the characteristics of these sources, together with references for the time-lag spectra we used. The data of NGC\,4593 are taken from \cite{Cackett18}, which includes data from the \textit{HST} and \textit{Swift} monitoring. We note that these results are in agreement with a previous analysis of the \textit{Swift} data by \cite{Mchardy18}. The  NGC\,5548, Mrk\,817, and Fairall\,9 data are taken from \cite{Fausnaugh16}, \cite{Kara21}, and \cite{Santisteban20}, respectively. The Eddington ratios, $\lambda_{\rm Edd}=L_{\rm bol}/L_{\rm Edd}$, are listed in the fourth column of Table\,\ref{tab:sources}, and are based on measurements reported in the aforementioned papers ($L_{\rm bol}$ and $L_{\rm Edd}$ are the bolometric and Eddington luminosities, respectively). We will accept those as indicators of the actual accretion rate of the source, normalised to the Eddington limit, although this assumption is not straight forward (see below). As for the X--ray properties (photon index and $2-10$\,keV luminosity) of the sources, we considered the values reported in K21a for NGC\,5548 and NGC\,4593. For Mrk\,817, we considered the values reported by \citet{Kara21} for the \textit{XMM-Newton} observation which coincides with a state where the source is close to its mean flux during the monitoring campaign. As for Fairall\,9, we followed the approach of K21a and we fitted the spectrum of the source by considering time intervals in which the source was close to its average flux during the monitoring campaign. The spectrum of the source was extracted using the automatic \textit{Swift}/XRT generator\footnote{\url{https://www.swift.ac.uk/user_objects/}} \citep{Evans09}. We fit the spectrum assuming a power-law model plus a reflection \citep[{\tt xillver;}][]{Garcia2013,Garcia16}, considering only Galactic absorption. The model can be written in the {\tt XSPEC} parlance as follows: 

$$ {\tt model = TBabs \times (powerlaw + xillver}). $$

\noindent We fixed the Galactic column density to $N_{\rm H} = 2.86\times 10^{20}\,\rm cm^{-2}$ \citep{HI4PI}, the inclination of {\tt XILLVER} to $20^\circ$, and the iron abundance to the solar value. The model resulted in a statistically good fit ($\rm\chi^2/dof = 218/197$) with a best-fit photon index $\Gamma = 1.90 \pm 0.03 $, an ionization parameter of the reflecting medium of $\log \left(\xi/\rm erg\,cm\,s^{-1}\right) = 1.3_{-0.3}^{+0.2}$, and an intrinsic X--ray luminosity of $\rm \log L_{2-10} = 43.99 \pm 0.01$. The values of the photon index and X--ray luminosity for all the four sources are reported in the last two columns of Table\,\ref{tab:sources}, and will be used in the rest of this analysis. 

We note that for all sources we used the centroid values of the ICCF as reported in the corresponding papers. For NGC\,4593 and Mrk\,817, we changed the reference wavelength to be the smallest one (1150\,\AA\ and 1180\,\AA, respectively) in order to avoid any contribution from the Balmer jump.


\section{Time-lags fitting}
\label{sec:lagfit}

X--ray to UV/optical time lags depend on many physical parameters of the system, such as the BH mass, spin, accretion rate, height of the corona, inclination, photon index, \ltransf, and $R_{\rm out}$. Furthermore, one may assume an X--ray source which is powered by the accretion process, or by an external source of power. Hence, it is not straight froward to fit the observed time lags in practice. We believe it is not possible to fit the time lag spectra by letting all parameters free to vary. In fact, some of the model parameters are degenerate (i.e., \ltransf, \mdot, and $h$ can all affect the time lags in similar ways), and in some cases, the number of parameters may even be larger than the number of points in the observed time lag spectra. We present below a possible approach in fitting the observed time lags in the four AGN in our sample. Throughout this analysis, we will assume that a part of the accretion power that is released below a certain radius is transferred to the corona (by an unknown mechanism), hence, we will assume a positive \ltransf.

\subsection{Fitting method}
\label{sec:fittingmethod}

\begin{table}
\caption{The sources in our sample. Luminosity distance, $D_{\rm L}$, BH mass, $M_{\rm BH}$, the Eddington ratio ($\lambda_{\rm Edd}$), the photon index ($\Gamma$), and the $2-10$\,keV luminosity obtained from the X--ray spectral analysis (see text for details).}
\label{tab:sources}
\begin{threeparttable}
\begin{tabular}{llllll} 
\hline \hline
Source	    &	 $D_{\rm L}^{^a}$	& $M_{\rm BH}^{^{b}}$    &	$\lambda_{\rm Edd}^{^{c}}$	&	  	 $\Gamma$ & $L_{\rm 2-10}^{^d}$ 	\\
	        & (Mpc)	        & $	  (10^7\,M_\odot$)       &		    &	&	($10^{43}\,\rm cgs$)             \\ \hline

NGC\,4593		        & 38.8	        & $ 0.76^{+ 0.16}_{-0.16}$  &	0.08	& 1.74	 & $0.6\pm 0.2$  \\ [0.1cm]
NGC\,5548	 	        & 80.1	        & $	7.0 ^{+2.0}_{-2.8}$           &	0.05	&  	1.70         &  $2.5 \pm 0.7$     \\ [0.1cm]
Fairall\,9		        & 209       & $	19.9_{-4.6}^{+3.9}$           &	0.03	&  	1.90         &  $9.7 \pm 1.9$        \\ [0.1cm] 
Mrk\,817		        & 138.7	        & $	3.8^{+0.6}_{-0.6} $           &	0.20	&  	1.90  &  $1.9 
\pm 1.3$           \\ [0.1cm]\hline

\end{tabular}
\begin{tablenotes}
\item[$^a$]  We computed $D_{\rm L}$ using the source redshift,  assuming a flat Universe with: $H_0 = 67.8\,\rm km\,s^{-1}\,Mpc^{-1}$, $\Omega_{\rm \Lambda} = 0.7$, and $\Omega_{\rm M} = 0.3$. 
\item[$^b$]$M_{\rm BH}$ are from ``The AGN Black Hole Mass Database'' \citep{Bentz15} except for NGC\,5548 we use the estimate by \citet{Horne2021}; we use the estimates when considering all emission lines.
\item[$^c$] References to the Eddington rations: \citet{Cackett18} for NGC\,4593, \citet{Fausnaugh16} for NGC\,5548, \citet{Vasudevan07} for Fairall\,9, and  \citet{Kara21} for Mrk\,817. 
\item[$^d$] The uncertainty on the $2-10$\,keV luminosity represent the scatter around the mean in the X--ray light curves of each of the sources.

\end{tablenotes}
\end{threeparttable}

\end{table}

In order to avoid the possible degeneracy between the various parameters and to speed up the fitting process, we proceeded as follows. For each of the sources, we fixed the BH mass and the photon index to the values given in Table\,\ref{tab:sources}. We also fixed the inclination to 45\degr. Then we considered three values of the BH spin (0., 0.7, and 0.998), three values of \ltransf\ $(0.1, 0,5, \,\rm and\, 0.9)$, and three values of \fcol\ (1, 1.7, and 2.5). For each combination of these values we fixed the height at 10 values between 3\,\rg\ and 96\,\rg\ fitting only for \mdot. This results in 270 different combinations of the four model parameters ($a^\ast$, $h$, \ltransf, \fcol), with \mdot\ being the only free parameter of the fit. While the data quality does not allow for a direct fitting to be applied, leaving all parameters free, our approach can be used to exclude specific parts of the parameter range, and to also constrain the parameters of interest when including independent information, as will be shown below..

We fitted the time lags using a combination of two functions from the {\tt scipy.optimize} library: {\tt curve\_fit}\footnote{\url{https://docs.scipy.org/doc/scipy/reference/generated/scipy.optimize.curve_fit.html}} and {\tt basinhopping}\footnote{\url{https://docs.scipy.org/doc/scipy/reference/generated/scipy.optimize.basinhopping.html}}. The first function solves a non-linear least square problem with bounds on the variables. This method was chosen in order to reduce the fitting time comparing to other fitting scheme in the same library. However, using this method does not avoid the risk of falling into a local minimum. To overcome this problem, we use the {\tt basinhopping} algorithm \citep{Wales97}, which first minimizes $\chi^2$, then randomly chooses a new starting point depending on the minimization result and launches a new minimization starting from this new point. After that, the results of both minimizations are compared based on their $\chi^2$ values. The best result is kept and the algorithm repeats the above starting from the point it has selected. The number of such cycles can be specified by the user as a parameter of this function. For our fitting tool, we modified the {\tt basinhopping} method from {\tt SciPy} to use {\tt curve\_fit} as the optimization method. We found that, since we are fitting for only one parameter, the fit converges to a point independent of the starting point after two {\tt basinhopping} cycles, which is assumed to be the global minimum.

\subsection{Results}
\label{sec:bestfitres}

\begin{figure}
    \centering
    \includegraphics[width=1\linewidth]{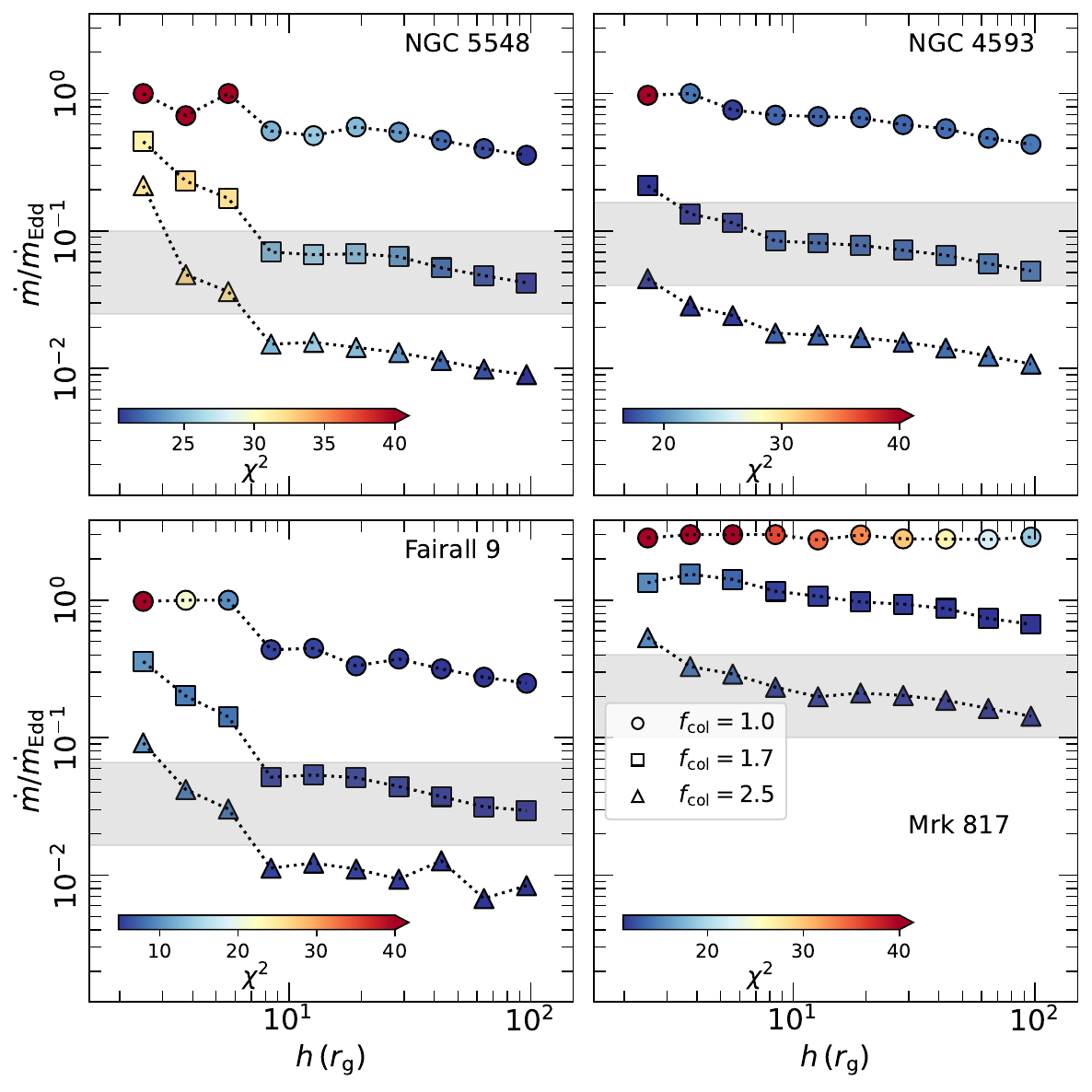}

\caption{Accretion rate versus coronal height for all of the sources assuming $a^\ast = 0$ and $\ltransf = 0.5$ for $\fcol
= 1, 1.7,$ and 2.5 (circles, squares, and triangles, respectively). The color bars correspond to the value of $\chi^2$ obtained for each fit (the degrees of freedom for NGC\,5548, NGC\,4593, Fairall\,9, and Mrk\,817 are 12, 14, 6, and 12, respectively). The shaded areas correspond to the uncertainty by a factor of two around the Eddington ratio obtained from the literature for each of the sources.}
\label{fig:mdot_height}
\end{figure}


We fit the time lag spectra of each of the sources as described above. We did not consider time-lags measurements between 2000-4000\AA\,  in NGC\,4593. According to \citep{Cackett18}, the time-lags spectrum of NGC\,4593 shows a clear excess around the 3646\AA\, Balmer jump, which could imply that diffuse emission from gas in  broad-line region (BLR) may contribute significantly to the observed time-lags. \cite{Edelson15} also noticed that the {\it U} band time-lag measurement of NGC 5548 was larger than the measurements in the surrounding bands, which could also be attributed to the Balmer continuum emission. Similar comments were made by \cite{Santisteban20} in the case of the Fairall\,9 time-lags observations.  For that reason, we did not consider the measurements in the {\it Swift/U} and ground based {\it u} bands in NGC\,5548 and Fairall\,9. We considered all time-lags measurements in the case of Mrk817.

For Fairall\,9 and NGC\,5548 we fixed the outer radius of the disc to $R_{\rm out}=5000\,\rg$. However, for NGC\,4593 and Mrk\,817, this value of $R_{\rm out}$ underestimated the lag at longer wavelength. This is due to the fact that these two sources have lower BH masses and higher accretion rates compared to the former two sources. Thus, their discs are hotter, and a larger value of $R_{\rm out}$ is needed to fit the time lag spectra. So we fixed it at 10000\,\rg\ for NGC\,4593 and Mrk\,817.


\begin{figure*}
    \centering
    \includegraphics[width=1\linewidth]{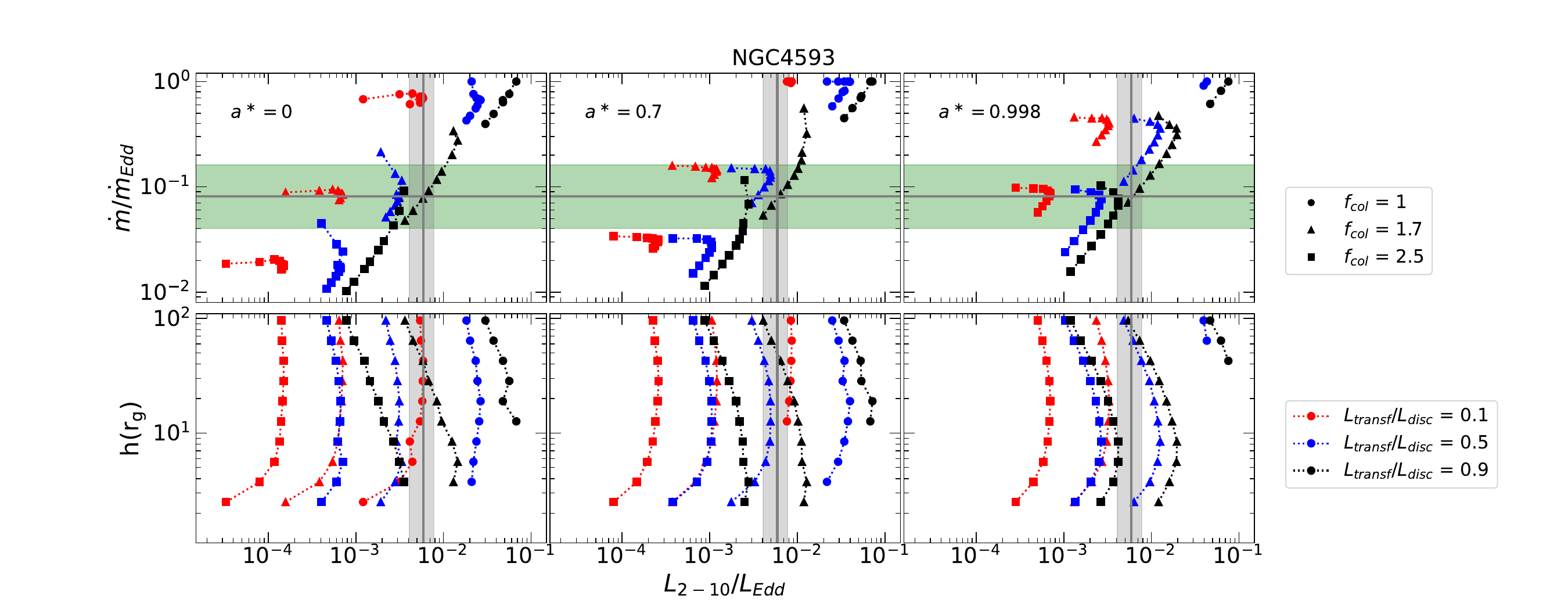}
    \includegraphics[width=1\linewidth]{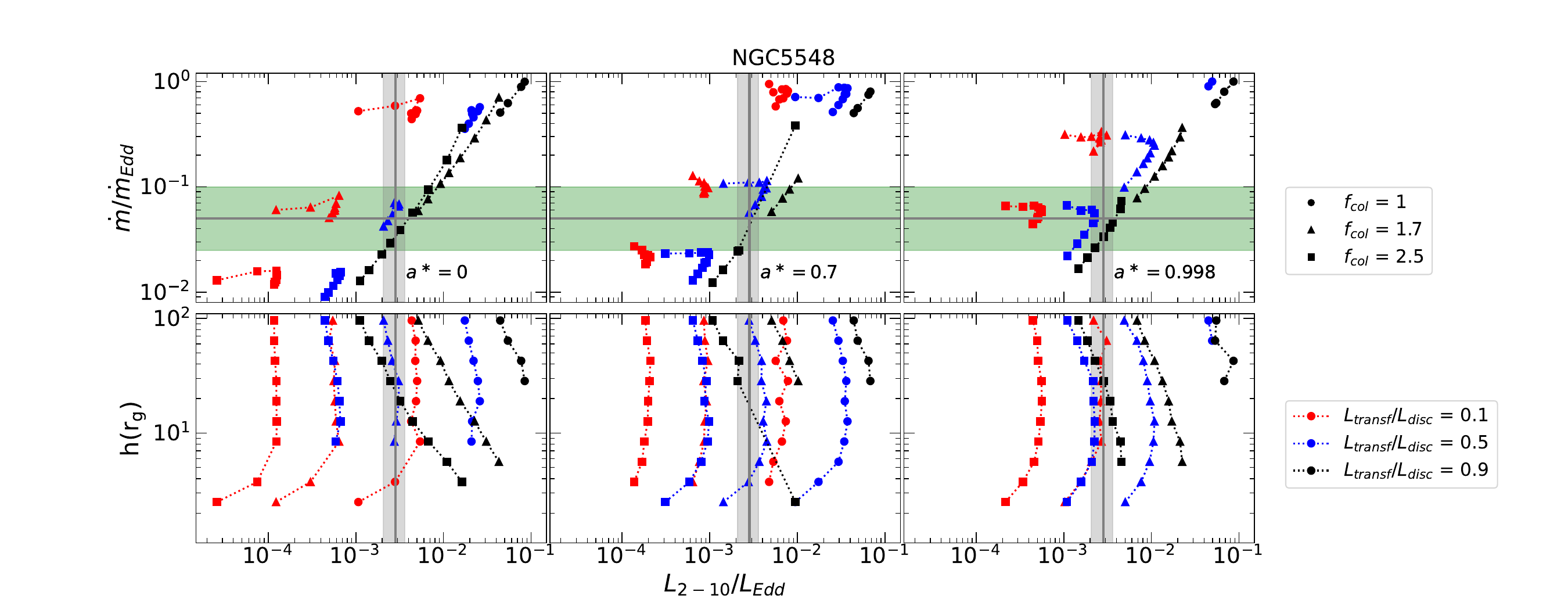}
    \includegraphics[width=1\linewidth]{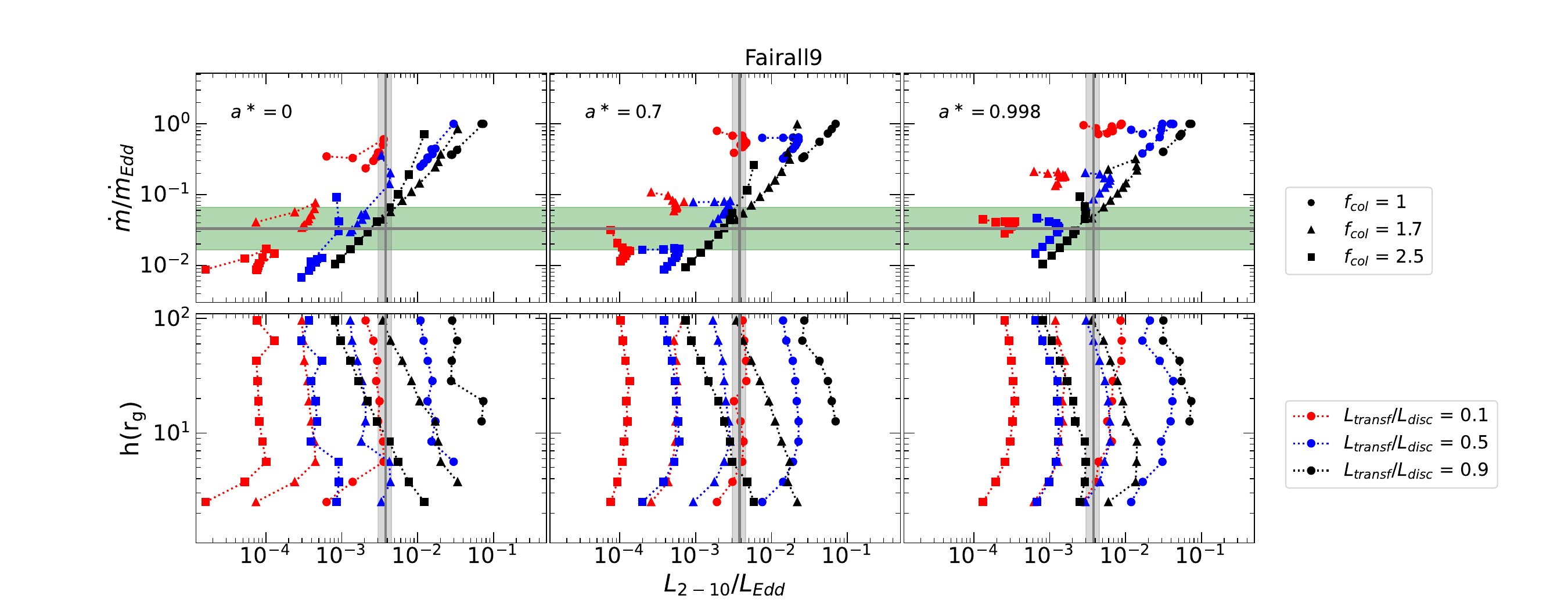}
\caption{Best-fit \mdot\ and $h$ versus \Lxfit\ (top and bottom rows, respectively) for NGC\,4593, NGC\,5548, and Fairall\,9 (top to bottom). The results are shown for $a^\ast = 0, 0.7,$ and 0.998 (left to right) for $\ltransf = 0.1, 0.5,$ and 0.9 (red, blue, and black, respectively) and $\fcol = 1, 1.7,$ and 2.5 (circles, triangles, and squares, respectively). The vertical lines and the shaded grey areas represent the average value of \Lxobs\ and the corresponding $1\sigma$ uncertainty. The horizontal lines and the shaded green areas represent the values of the Eddington ratio obtained from the literature and the corresponding factor of 2 uncertainty, respectively (See text for details).}
\label{fig:lumxray_all}
\end{figure*}

\begin{figure*}
    \centering
    \includegraphics[width=1\linewidth]{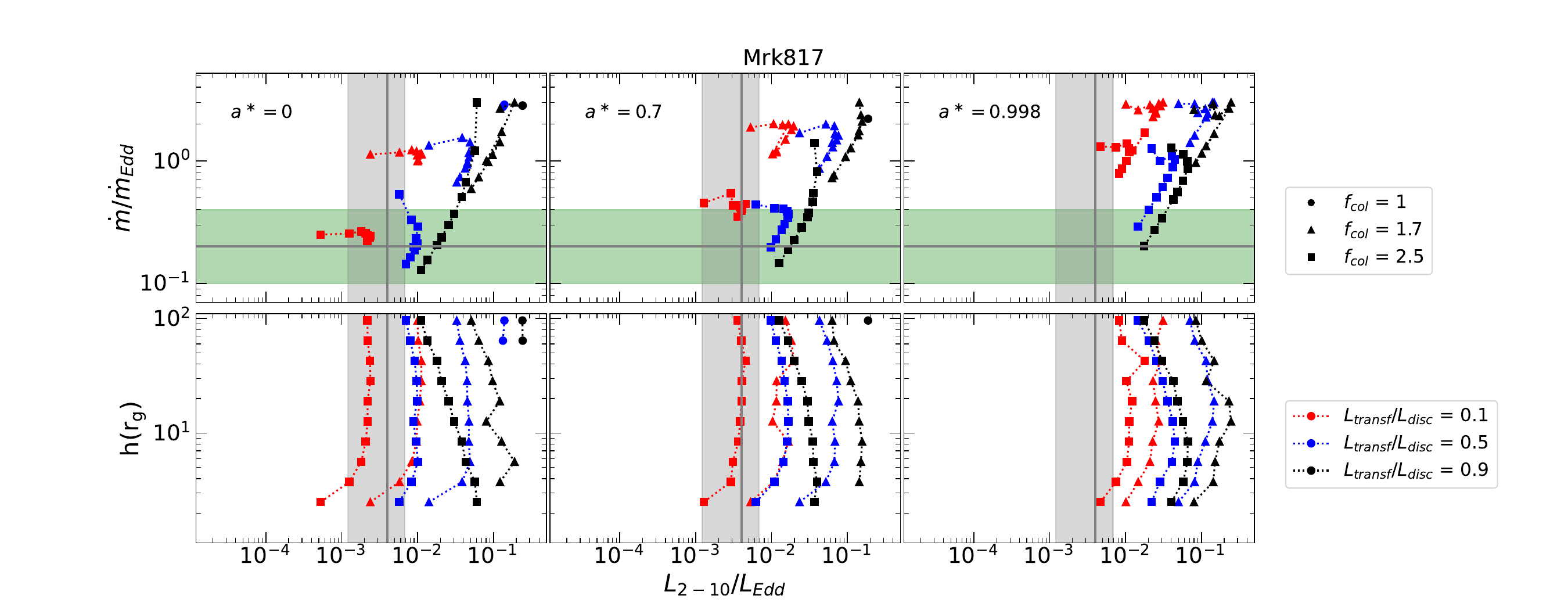}
\contcaption{Best-fit results for Mrk\,817.}
\label{fig:lumxray_mrk817}
\end{figure*}


Figures\,\ref{fig:timelag_NGC4593}-\ref{fig:timelag_Mrk817} of the appendix show the time lag spectra fitted using the various combinations of spin, height, \fcol, and \ltransf\ considered in this work by letting \mdot\ as a free parameter. As it can be seen from these figures, a wide range of parameters can provide a good fit to the data (for different mdot values), for all of the sources. However, not all parameters are accepted. For example, the case $a^\ast = 0$, $\ltransf = 0.9$, and $\fcol = 1$ in Fig.\,\ref{fig:timelag_NGC5548} (bottom panel) clearly shows that the low height values do not provide a good fit to the time-lags spectra. 

The range of the best-fit \mdot\ and height depends strongly on the assumed combination of ($a^\ast$, \ltransf, \fcol). This is illustrated in Fig.\,\ref{fig:mdot_height}, where we show the best-fit \mdot\ as a function of the corona height for the various \fcol\ values we consider, assuming $a^\ast = 0$ and $\ltransf = 0.5$. The colour code in this figure shows the $\chi^2$ value obtained in each best-fit (for all sources). The plots in this figure clearly show that various combinations of \mdot and X--ray source height can fit the observed time-lags, in all sources, equally well.

Additional constraints can be put on the resulting best-fit parameters by checking which values of the best-fitting \mdot\ are in agreement with the $\lambda_{\rm Edd}$ values listed in Table\,\ref{tab:sources}. 
As we mentioned above, we use $\lambda_{\rm Edd}$ as a measure of the accretion rate in each source, however this is highly uncertain. For example, the bolometric luminosity is usually based on measuring the observed luminosity in a given spectral band and then applying a bolometric correction factor. However, the observed luminosity may not be representative of the mean luminosity in this band, and the correction factor is quite uncertain. Furthermore, converting from $L_{\rm bol}/L_{\rm Edd}$ to \mdot\ requires a knowledge of the inclination of the system (which we do not know). Thus, we consider a conservative uncertainty on the observed $L_{\rm bol}/L_{\rm Edd}$ by a factor of $\sim 2$, to account for all the aforementioned factors. 

The horizontal grey shaded areas in Fig.\,\ref{fig:mdot_height} indicate the $\lambda_{\rm Edd}$ values listed in Table\,\ref{tab:sources} together with the assumed uncertainty, as explained above. For the specific combination of $a^\ast = 0$ and $\ltransf = 0.5$, Fig.\,\ref{fig:mdot_height} shows that $\fcol = 1$ is not an accepted value, as all best-fitting values of \mdot\ are above the grey area, for all sources. On the other hand, there are quite a few heights for which the best-fit \mdot\ lies within the grey area when $\fcol=1.7$ and 2.5. In some cases, the range of the heights which results in \mdot\ within the grey area is relatively narrow. See for example the bottom left panel in Fig.\,\ref{fig:mdot_height}, for Fairall 9. The range of the accepted heights is quite narrow for $\fcol=1.7$, while $\fcol=2.5$ puts a strong low limit on the corona height at 10\,\rg. We conclude that, although many model parameters can result in model time-lags which can fit the observations well, when we consider the (assumed) accretion rate for each source, we can still rule out some model parameter combinations.

An additional selection can be performed to reduce further the best-fitting parameter space. For a certain set of parameters, the code computes a posteriori the expected observed $2-10$\,keV luminosity. This quantity can be directly compared to the observed, average $2-10$\,keV luminosity, as listed in Table\,\ref{tab:sources}. Figure\,\ref{fig:lumxray_all} shows the best-fit \mdot\ and height values (upper and bottom panels, for each source), as a function of the model $L_{2-10}$ in units of the Eddington luminosity $L_{\rm Edd}$ (hereafter \Lxfit). In this figure, we show only the mdot and heights that lead to a statistically acceptable fit of the time lag spectra (i.e.\, $p_{\rm null} > 0.01$). Each of the columns in this figure corresponds to a fixed spin value. We show in different colors the results for various \ltransf, and in different symbols the results for various \fcol. The vertical lines indicate the observed $2-10$\,keV luminosity divided by $L_{\rm Edd}$ (\Lxobs). The vertical grey shaded area indicates the $1\sigma$ scatter around the mean as estimated from the \textit{Swift}/XRT light curves. The horizontal solid lines indicate the $\lambda_{\rm Edd}$ values listed in Table\,\ref{tab:sources}, while the green shaded area shows the uncertainty associated with it (as explained above). 

This figure shows clearly that, when we consider the additional constraints of the observed X--ray luminosity, then the accepted parameter space of the best-fit \mdot\ and height values is further reduced. For example, the left-hand side, top panels in Figure\,\ref{fig:lumxray_all} show the best-fit \mdot\ and height values in the case of NGC\,4593, when we assume a BH with zero spin. The uppermost panel indicates that only the model parameters indicated by the black triangles can fit the time lags {\it and}, at the same time, can also result in accretion rates {\it and} $2-10$\,keV band luminosity which are in agreement with the observations. Triangles imply an \fcol\ of 1.7, while the black color indicates an \ltransf\ of 0.9. The panel just below the uppermost left panel, show that only heights of $\sim 20-70\,\rg$ are consistent with both the \mdot\ and X--ray luminosity constraints. The red circles in the same panel indicate that the combination of $\fcol=1$, $\ltransf=0.1$ and a source height above $\sim 5\, \rg$ could also fit the observed time lags well and predict the observed X--ray luminosity, {\it but} the necessary \mdot\ value would be $\sim 8$ times larger than the $\lambda_{\rm Edd}$ value listed in Table\,\ref{tab:sources}, so we do not consider this as a probable combination of model parameters for this source.  

Following the same procedure, we selected the best-fit model parameters that can provide a good fit to the observed time lags and, at the same time, can also predict an accretion rate and $2-10$\,keV luminosity which are consistent with $\lambda_{\rm Edd}$ and the observed luminosity values listed in Table\,\ref{tab:sources}, within their uncertainties (defined as explained above).  Our final, best-fit model parameters for the four sources are listed in Table\,\ref{tab:bestift}. The last column in this Table lists the best-fit $\chi^2$ values together with the number of degrees of freedom (dof). All models listed in this table provide an acceptable fit to the data. For example, even in the case of the $\ltransf=0.9$, $a^{\ast}=0$, $\fcol=2.5$ model fit to the NGC\,5548 data, which resulted in the largest $\rm \chi^2/dof$ ratio, the null hypothesis probability is 0.025, i.e., larger than 0.01).

\begin{table}
    \caption{Range of the best-fit parameters obtained by fitting the time-lag spectra of each of the four sources.}
    \label{tab:bestift}
    \begin{tabular}{lllll}
    \hline \hline
    $\ltransf$ & $a^\ast$ & $\fcol$ & $h\, (\rg)$ & $\rm \chi^2_{min}/ dof $ \\ \hline
				\multicolumn{5}{c}{NGC\,4593}			\\[0.2cm]
$0.5$& $0.7$	&	$1.7$	&	$[5.6; 42.7]$ & $18.6/14$	\\
 & $ 0.998$	&	$1.7$	&	$\geq 64 $	& $19.9/14$ \\
	\\

$0.9$	& $ 0$	&	$1.7$	&	$[28.5; 64]$	& $\rm  16.8/14$\\
& $0.7$	&	$ 1.7$	&	$\geq 42.7$	& $\rm 18.4/14$ \\
 & $0.998$	&	$1.7$	&	$\geq 64 $	& $20.3/14$\\
	& $0.998$ &		$2.5$	&	$[5.6; 8.4]$	& $18.5/14$\\ \hline
				\multicolumn{5}{c}{NGC\,5548}			\\[0.2cm]
$0.5$	& $0$	&	$1.7$	&	$\geq 8.4 $	& $20.4/12$\\
& $0.7$	&	$1.7$	&	$\geq 64 $	& $ 19.7/12$\\
& $0.998$	&	$2.5$	&	$[5.6; 28.5]$	& $20.3/12$\\
							\\
$0.9$	& $ 0$	&	$2.5$	&	$[18.9; 28.5] $	& $ 23.4/12$\\ 
& $ 0.998$	&	$2.5$	&	$[18.9; 42.7] $	& $20.5/12$\\ \hline
				\multicolumn{5}{c}{Fairall\,9}			\\[0.2cm]
$ 0.9$	& $0$	&	$1.7$	&	$\geq 64 $	& $5.4/6$\\
	& $0$ &	$2.5$	&	$8.4 $ & $ 5.5/6$	\\
& $0.7$	&	$1.7$	&	$ \geq 64 $ & $5.1/6$	\\
& $0.998$	&	$1.7$	&	$96 $ & $4.7/6$	\\ \hline
				\multicolumn{5}{c}{Mrk\,817}			\\[0.2cm]
$0.1$	& $ 0$	&	$ 2.5$	&	$ \geq 4 $	& $11.2/12$ \\
& $ 0.7$	&	$ 2.5$	&	$ \geq 64 $ & $11.2/12$ \\ \hline
    \end{tabular}
\end{table}

If we accept the $\lambda_{\rm Edd}$ values listed in Table\,\ref{tab:sources} as reliable estimates of the accretion rate in these objects, our results indicate that \fcol\ cannot be equal to one, i.e., the emission from the accretion disc in these objects is not simply equal to the flux emitted by multi-temperature black-bodies, where the temperature depends on radius according to the \citet{Novikov73} and \citet{Shakura73} prescriptions. In addition, our results show that, if we assume that the X--ray source is powered by the accretion process, then our results indicate that a fraction of the accretion power released in the accretion disc that is larger than 50 per cent should be transferred to the X--ray corona. The exception is Mrk\,817, which is the object with the highest accretion rate among the four sources. Our results indicate that the X--ray corona luminosity should be smaller than 50 per cent of the accretion power in this source. This source is also the only one for which we cannot find a good fit to the data for $a^\ast=0.998$. All spins we consider can provide good fits to the data in the other three sources. Furthermore, the accepted range of the X--ray source height is rather broad. In many cases we find solutions where the source height is large (this is mainly the case with Fairall\,9), but we also get solutions where the X--ray source height is as small as $\sim 5-6\,\rg$. 

NGC\,5548 and NGC\,4593 are the only sources in common with K21a. For NGC\,5548, K21a reported $h \in [23\,\rg, 58\,\rg]\, ([33\,\rg, 78\,\rg])$ and $\mdot \leq 0.008\, (0.05)$ for $a^\ast = 0\, (a^\ast = 1)$, at the $1\sigma$ confidence level. Considering $\fcol = 2.5$ (comparable to the assumptions of K21b), we find a good agreement with the results of K21a for high spin values. We find a larger value of \mdot, while the ranges of height are in agreement, for the non-spinning BH case. The larger  \mdot\ value is due to the fact that we consider the case of an X--ray source powered via accretion. This results ino smaller time lags compared to an X--ray source powered externally, which was the case in K21a. To compensate for this, the  model fita the data  but with a larger  \mdot. As for NGC\,4593, K21a report  $h \leq 25\,\rg\, (23\,\rg)$ and $\mdot \in [0.006, 0.016]\, ([0.06, 0.22])$ for $a^\ast = 0\, (a^\ast = 1)$, at the $1\sigma$ confidence level. For the non-spinning case, we do not find any solution that fits the time lag spectra. This is comparable to the results of K21a, as the values of \mdot\ are smaller than the observed Eddington ratio. The results for a maximally spinning black hole are in agreement with K21a. We note that for both sources, the 1$\sigma$  parameter range from the current fit is significantly smaller when compared to K21a.

\section{Discussion and Conclusions}
\label{sec:conclusions}

We present a new code which can be used to fit the time lag spectra that has been observed the last few years in many AGN , under the assumption of disc thermal reverberation, due to X--ray illumination. Our work extends the work presented by K21b. These authors presented analytic functions for the time lags in the case when the source that powers the X--ray corona is not associated with the accretion power and the BH spin is either zero or one. They also assumed a color correction factor of $\fcol=2.4$, and an infinite outer disc radius. The new code is based on the work of K21b, as it assumes a point-like X--ray source illuminating an NT accretion disc in the lamp-post geometry. However, we also take into account the recent work of D22. Consequently, the new code offers significant improvements when compared with the analytic functions of K21b in many ways: a) it can be used both in the case when the X--ray source is powered by a source which is not associated with the accretion process (as in K21b) but also in the case when the X--ray luminosity is a fraction of the accretion process (which is, somehow, transferred to the X--ray corona) b) it can be used to fit the data for any BH spin, from zero to 0.998, c) any \fcol\ value, and d) any outer disc radius. We note that, the analytic functions of K21b, as well as the new code, can be used to fit time-lags by keeping the BH mass as a free parameter. As we have already argued, one should try to minimize the number of free parameters when fitting the time-lags, however, recent work by \cite{Pozonunez2019} suggests that black hole mass measurements in AGN may be underestimated  due to the unknown BLR geometry. In this case, one can leave the BH mass as a free parameter when fitting the time-lags in order to investigate this possibility. 

We studied in detail the dependence of the time-lags on \ltransf\ in the case when the X--ray source is powered by an external source and in the case when it is powered by the accretion process. We find that, for the same \ltransf, the time-lags are smaller in the latter case. This is because, in this case, the (relative) contribution of the inner disc emission to the observed flux in each optical/UV band increases with respect to the case of an externally powered X--ray corona. As for \fcol, we find strong effects on the time lags that are similar to the effects of the accretion rate. The time-lags increase with larger \fcol, just like the time-lags increase with increasing accretion rate.  Therefore, \fcol\ and \mdot\ (at least) should be degenerate, when it comes to the determination of either parameter from fits to the observed time-lags. It is also worth noting that in this work we use the values of Eddington ratio reported in the literature, which do not take into consideration the effect of intrinsic reddening that may be important \citep[see e.g.,][]{Gaskell2023}. If intrinsic reddening is significant, this will affect the value of the observed Eddington ratio. This should not alter the shape of the time lags, however it may affect the values of the best-fit parameters.

We used the code to fit the observed time-lags in four AGN. We chose these objects mainly because their time lags have been determined in many wavebands, from the far-UV up to $\sim 8000-9000$\,\AA, thus they are the best time lag spectra to test theoretical models. To fit the data, we fix the BH mass and outer disc radius, and then we consider a large number of \fcol, \ltransf, and spin values, and we fit the data by letting just \mdot\ and the X--ray source height to be free parameters. This approach is necessary in order to fit the data because, as we have already mentioned, many model parameters affect the time lags in similar ways, hence introducing various degrees of degeneracy between the parameters. We find that the time lags of all sources are well fitted by the model for a large range of model parameters (see for example the best-fit models plotted in Figs.\,\ref{fig:timelag_NGC4593}-\ref{fig:timelag_Mrk817} in appendix). By introducing 
further observational constraints such as the observed $\lambda_{\rm Edd}$ and the observed $2-10\,\rm keV$ luminosity, we are able to reduce the range of parameters that can explain the time lag spectra and be consistent with the other constrains as well is reduced.

For all sources, we find values of \fcol\ that are greater than\,1. We emphasize that this result depends on whether the observed $\lambda_{\rm Edd} $values listed in Table\,\ref{tab:sources} are indeed representative of the accretion rate in these objects or not. If the intrinsic accretion rate is larger in these objects, then \fcol\ could be closer to unity. Our results are in agreement with previous works such as \citet{Shimura95, Ross92, Done2012, Davis19}, who indicated the need for modifications to the standard blackbody accretion disc model, in the form of a colour temperature corrected blackbody. In particular, \citet{Davis19} found a moderate variation of \fcol\, between $1.4 - 2$, for accretion rates between 0.01 and 1 of the Eddington rate. Their Eq.\,(10) presents an analytic approximation of \fcol\ as a function of the BH mass and accretion rate for $a^\ast = 0 $ and 0.9. We used this equation and we found that \fcol\, factors of $\sim 1.6(1.7)$, 1.55(1.65), 1.4(1.5) and 1.73(1.83) for NGC\,4593, NGC\,5548, Fairall\,9 and Mrk\,817, respectively, in the case of $a^\ast=0(0.9)$, when assuming that $\alpha=0.1$. Our results, based on time-lag modelling, predict slightly larger values of \fcol, but we did not investigate a dense range of \fcol\ values in our work. We suspect that models with \fcol\ values equal to the ones reported above will  almost certainly give good fits to the observed time-lags. However, \citet{Davis19} considered a non X--ray illuminated accretion disc. In fact, for X--ray illuminated discs, \fcol\ may be slightly different than the ones reported by \citet{Davis19}. Given this uncertainty, we believe our measurements are in very good agreement with the predictions of \citet{Davis19}, and this result indicates that colour correction factors may be necessary when fitting broadband AGN SEDs, as well as timing results.

We suggest to always use {\tt KYNXiltr} when fitting the observed time-lag spectra, as it can cover both scenarios of the externally and the internally heated X--ray corona. In fact, in this way, we may get indirect evidence regarding the source of power that heats the X--ray corona. For example, if the internally powered X--ray corona models do not fit well the time-lag spectra, even for the maximum allowed value of $\ltransf=0.9$, this could be an indication that the X--ray corona is powered by other mechanisms (most probably associated with transfer of power from the vicinity of the BH), as long as the model can fit the data in the case of the externally heated corona with a larger \ltransf. The analytic expressions of K21b are valid in the case of an externally heated corona, as long as $a^\ast=0$ or 1, and the outer disc radius is larger than 10000\,\rg, and $\fcol=2.4$. The restriction of the two spin values may not be very strong, as it seems rather unlikely that it would be possible to derive a spin parameter estimate with a small error with the currently observed time-lag spectra, given the large number of physical parameters and their inter-dependencies. We provide updated versions of the analytic expressions in Appendix\,\ref{sec:analytic}, which take into account the model dependence on \fcol. We also discuss ways  one could get a rough estimate of time lags in the case of internally powered X--ray coron\ae, using the analytic time-lag equations of K21b. 

When fitting the time-lags, we omitted data points in the U-band (NGC\,5548 and Fairall\,9) while in the case of NGC\,4593, between $2000-4000$\,\AA. This is because the time-lags measurements in these wavelengths may be affected by line and continuum emission from gas in the BLR. This issue was already noticed by many authors in the past. Recently,  \cite{Netzer2022} suggested that the total lag-spectrum, and its normalization, could be due to diffuse emission from radiation pressure supported clouds in a BLR with a covering factor of about 0.2. Their results were based on the assumption of X-ray disc thermal reverberation time-lags which are significantly different than the ones we present here. For example, the continuum time-lags shown by the dashed line in Fig.\,1 of \citet{Netzer2022} are very different than our best-fit models presented in Figs.\,\ref{fig:timelag_NGC4593}-\ref{fig:timelag_Mrk817}. 

Our model fits show that the majority of the observed time-lags can be fully explained when self-consistently modelling the X--ray irradiation of an NT accretion disc, for reasonable physical parameters. We fit the Mrk\,817 time-lags spectrum very well without omitting any points in the U-band. All the observed time lags in this object can be fully accounted by X-ray, thermal reverberation. In the case of Fairall\,9, the best-list models listed in Table \ref{tab:bestift} can provide an acceptable fit to the full time-lags spectrum even if we consider the U-band measurements ($\chi^2$ increases from $5-5.5$ to $15-15.5$/8 dof, for all the models listed in Table \ref{tab:bestift}, which implies $p_{null}\sim 0.05-0.06$). The U-band bump is more pronounced in the case of NGC\,5548, where $p_{\rm null}$ becomes less than 0.01 when we consider the U-band time-lags as well, for some of the best-fit models listed in Table \ref{tab:bestift}.  However, we do not detect the $5000 - 10,000$\,\AA\, feature (close to the wavelength of the Paschen jump at 8200\,\AA), as suggested by \citet{Netzer2022}, in the best fit residuals of NGC 5548 and Fairall 9. 

The X-ray reverberation time-lags cannot account for the U-band bump in the NGC\,4593 time-lags. A bump at around $6000-8000$\,\AA\ is also observed in this object, although it is not as pronounced as the one in the U-band. We therefore conclude that, at least in some AGN, the U-band time lags cannot be fully explained if we assume only X--rays illumination of the disc. It is unclear why the time-lags due to diffuse emission in the BLR may be present in one object and not in others. It may depend on the geometry of the BLR, the way the central source illuminates the clouds etc. Investigating this issue is not straightforward. In any case, we plan to investigate in the future whether we can update our code by including the contribution of the time-lags from the diffuse BLR emission.

Our work shows that considering only time lags is not enough to directly constrain all the different parameters of the corona/disc configuration. In order to entirely lift any degeneracy between the various model parameters, and find a unique set of parameters that can explain the observed data, it is also required to fit the observed broad-band X--ray to UV/Optical SED  \citep{Dovciak22} and the power spectra \citep{Panagiotou20, Panagiotou22}. This will be invoked in follow-up publications.

\section*{Acknowledgements}

We thank the anonymous referee for their helpful comments. ESK and LR acknowledges financial support from the Centre National d’Etudes Spatiales (CNES) from which part of this work was completed. IEP acknowledges support from the
European Union’s Horizon 2020 Programme under the AHEAD2020 project (grant agreement n. 871158). This work made use of data supplied by the UK \textit{Swift} Science Data Centre at the University of Leicester. This work makes use of Matplotlib \citep{Matplotlib}, NumPy \citep{Numpy}, and SciPy \citep{Scipy}.

\section*{Data Availability}

The data for the time-lag spectra are all published in previous works cited in Table\,\ref{tab:sources}. The code is publicly available at \url{https://projects.asu.cas.cz/dovciak/kynxiltr}.


\bibliographystyle{mnras}
\bibliography{reference} 

\begin{thebibliography}{}
\makeatletter
\relax
\def\mn@urlcharsother{\let\do\@makeother \do\$\do\&\do\#\do\^\do\_\do\%\do\~}
\def\mn@doi{\begingroup\mn@urlcharsother \@ifnextchar [ {\mn@doi@}
  {\mn@doi@[]}}
\def\mn@doi@[#1]#2{\def\@tempa{#1}\ifx\@tempa\@empty \href
  {http://dx.doi.org/#2} {doi:#2}\else \href {http://dx.doi.org/#2} {#1}\fi
  \endgroup}
\def\mn@eprint#1#2{\mn@eprint@#1:#2::\@nil}
\def\mn@eprint@arXiv#1{\href {http://arxiv.org/abs/#1} {{\tt arXiv:#1}}}
\def\mn@eprint@dblp#1{\href {http://dblp.uni-trier.de/rec/bibtex/#1.xml}
  {dblp:#1}}
\def\mn@eprint@#1:#2:#3:#4\@nil{\def\@tempa {#1}\def\@tempb {#2}\def\@tempc
  {#3}\ifx \@tempc \@empty \let \@tempc \@tempb \let \@tempb \@tempa \fi \ifx
  \@tempb \@empty \def\@tempb {arXiv}\fi \@ifundefined
  {mn@eprint@\@tempb}{\@tempb:\@tempc}{\expandafter \expandafter \csname
  mn@eprint@\@tempb\endcsname \expandafter{\@tempc}}}

\bibitem[\protect\citeauthoryear{{Bentz} \& {Katz}}{{Bentz} \&
  {Katz}}{2015}]{Bentz15}
{Bentz} M.~C.,  {Katz} S.,  2015, \mn@doi [\pasp] {10.1086/679601}, \href
  {http://adsabs.harvard.edu/abs/2015PASP..127...67B} {127, 67}

\bibitem[\protect\citeauthoryear{{Cackett}, {Horne}  \& {Winkler}}{{Cackett}
  et~al.}{2007}]{Cackett07}
{Cackett} E.~M.,  {Horne} K.,   {Winkler} H.,  2007, \mn@doi [\mnras]
  {10.1111/j.1365-2966.2007.12098.x}, \href
  {http://adsabs.harvard.edu/abs/2007MNRAS.380..669C} {380, 669}

\bibitem[\protect\citeauthoryear{{Cackett}, {Chiang}, {McHardy}, {Edelson},
  {Goad}, {Horne}  \& {Korista}}{{Cackett} et~al.}{2018}]{Cackett18}
{Cackett} E.~M.,  {Chiang} C.-Y.,  {McHardy} I.,  {Edelson} R.,  {Goad} M.~R.,
  {Horne} K.,   {Korista} K.~T.,  2018, \mn@doi [\apj]
  {10.3847/1538-4357/aab4f7}, \href
  {http://adsabs.harvard.edu/abs/2018ApJ...857...53C} {857, 53}

\bibitem[\protect\citeauthoryear{{Cackett} et~al.,}{{Cackett}
  et~al.}{2020}]{Cackett20}
{Cackett} E.~M.,  et~al., 2020, \mn@doi [\apj] {10.3847/1538-4357/ab91b5},
  \href {https://ui.adsabs.harvard.edu/abs/2020ApJ...896....1C} {896, 1}

\bibitem[\protect\citeauthoryear{{Davis} \& {El-Abd}}{{Davis} \&
  {El-Abd}}{2019}]{Davis19}
{Davis} S.~W.,  {El-Abd} S.,  2019, \mn@doi [\apj] {10.3847/1538-4357/ab05c5},
  \href {https://ui.adsabs.harvard.edu/abs/2019ApJ...874...23D} {874, 23}

\bibitem[\protect\citeauthoryear{Done, Davis, Jin, Blaes  \& Ward}{Done
  et~al.}{2012}]{Done2012}
Done C.,  Davis S.~W.,  Jin C.,  Blaes O.,   Ward M.,  2012, \mn@doi [\mnras]
  {10.1111/j.1365-2966.2011.19779.x}, 420, 1848

\bibitem[\protect\citeauthoryear{{Donnan} et~al.,}{{Donnan}
  et~al.}{2023}]{Donnan23}
{Donnan} F.~R.,  et~al., 2023, \mn@doi [arXiv e-prints]
  {10.48550/arXiv.2302.09370}, \href
  {https://ui.adsabs.harvard.edu/abs/2023arXiv230209370D} {p. arXiv:2302.09370}

\bibitem[\protect\citeauthoryear{{Dov{\v{c}}iak}, {Papadakis}, {Kammoun}  \&
  {Zhang}}{{Dov{\v{c}}iak} et~al.}{2022}]{Dovciak22}
{Dov{\v{c}}iak} M.,  {Papadakis} I.~E.,  {Kammoun} E.~S.,   {Zhang} W.,  2022,
  \mn@doi [\aap] {10.1051/0004-6361/202142358}, \href
  {https://ui.adsabs.harvard.edu/abs/2022A&A...661A.135D} {661, A135}

\bibitem[\protect\citeauthoryear{{Edelson} et~al.,}{{Edelson}
  et~al.}{2015}]{Edelson15}
{Edelson} R.,  et~al., 2015, \mn@doi [\apj] {10.1088/0004-637X/806/1/129},
  \href {http://adsabs.harvard.edu/abs/2015ApJ...806..129E} {806, 129}

\bibitem[\protect\citeauthoryear{{Edelson} et~al.,}{{Edelson}
  et~al.}{2019}]{Edelson19}
{Edelson} R.,  et~al., 2019, \mn@doi [\apj] {10.3847/1538-4357/aaf3b4}, \href
  {http://adsabs.harvard.edu/abs/2019ApJ...870..123E} {870, 123}

\bibitem[\protect\citeauthoryear{{Evans} et~al.,}{{Evans}
  et~al.}{2009}]{Evans09}
{Evans} P.~A.,  et~al., 2009, \mn@doi [\mnras]
  {10.1111/j.1365-2966.2009.14913.x}, \href
  {https://ui.adsabs.harvard.edu/abs/2009MNRAS.397.1177E} {397, 1177}

\bibitem[\protect\citeauthoryear{{Fausnaugh} et~al.,}{{Fausnaugh}
  et~al.}{2016}]{Fausnaugh16}
{Fausnaugh} M.~M.,  et~al., 2016, \mn@doi [\apj] {10.3847/0004-637X/821/1/56},
  \href {http://adsabs.harvard.edu/abs/2016ApJ...821...56F} {821, 56}

\bibitem[\protect\citeauthoryear{Garc{\'{i}}a, Dauser, Reynolds, Kallman,
  McClintock, Wilms  \& Eikmann}{Garc{\'{i}}a et~al.}{2013}]{Garcia2013}
Garc{\'{i}}a J.,  Dauser T.,  Reynolds C.,  Kallman T.~R.,  McClintock J.~E.,
  Wilms J.,   Eikmann W.,  2013, \mn@doi [\apj] {10.1088/0004-637X/768/2/146},
  768, 146

\bibitem[\protect\citeauthoryear{{Garc{\'{\i}}a}, {Fabian}, {Kallman},
  {Dauser}, {Parker}, {McClintock}, {Steiner}  \& {Wilms}}{{Garc{\'{\i}}a}
  et~al.}{2016}]{Garcia16}
{Garc{\'{\i}}a} J.~A.,  {Fabian} A.~C.,  {Kallman} T.~R.,  {Dauser} T.,
  {Parker} M.~L.,  {McClintock} J.~E.,  {Steiner} J.~F.,   {Wilms} J.,  2016,
  \mn@doi [\mnras] {10.1093/mnras/stw1696}, \href
  {http://adsabs.harvard.edu/abs/2016MNRAS.462..751G} {462, 751}

\bibitem[\protect\citeauthoryear{{Gaskell}, {Anderson}, {Birmingham}  \&
  {Ghosh}}{{Gaskell} et~al.}{2023}]{Gaskell2023}
{Gaskell} C.~M.,  {Anderson} F.~C.,  {Birmingham} S.~{\'A}.,   {Ghosh} S.,
  2023, \mn@doi [\mnras] {10.1093/mnras/stac3333}, \href
  {https://ui.adsabs.harvard.edu/abs/2023MNRAS.519.4082G} {519, 4082}

\bibitem[\protect\citeauthoryear{{George} \& {Fabian}}{{George} \&
  {Fabian}}{1991}]{George91}
{George} I.~M.,  {Fabian} A.~C.,  1991, \mn@doi [\mnras]
  {10.1093/mnras/249.2.352}, \href
  {https://ui.adsabs.harvard.edu/abs/1991MNRAS.249..352G} {249, 352}

\bibitem[\protect\citeauthoryear{{HI4PI Collaboration} et~al.,}{{HI4PI
  Collaboration} et~al.}{2016}]{HI4PI}
{HI4PI Collaboration} et~al., 2016, \mn@doi [\aap]
  {10.1051/0004-6361/201629178}, \href
  {https://ui.adsabs.harvard.edu/abs/2016A&A...594A.116H} {594, A116}

\bibitem[\protect\citeauthoryear{{Haardt}}{{Haardt}}{1993}]{Haardt93}
{Haardt} F.,  1993, \mn@doi [\apj] {10.1086/173036}, \href
  {https://ui.adsabs.harvard.edu/abs/1993ApJ...413..680H} {413, 680}

\bibitem[\protect\citeauthoryear{{Harris} et~al.,}{{Harris}
  et~al.}{2020}]{Numpy}
{Harris} C.~R.,  et~al., 2020, \mn@doi [\nat] {10.1038/s41586-020-2649-2},
  \href {https://ui.adsabs.harvard.edu/abs/2020Natur.585..357H} {585, 357}

\bibitem[\protect\citeauthoryear{{Hern{\'a}ndez Santisteban}
  et~al.,}{{Hern{\'a}ndez Santisteban} et~al.}{2020}]{Santisteban20}
{Hern{\'a}ndez Santisteban} J.~V.,  et~al., 2020, \mn@doi [\mnras]
  {10.1093/mnras/staa2365}, \href
  {https://ui.adsabs.harvard.edu/abs/2020MNRAS.498.5399H} {498, 5399}

\bibitem[\protect\citeauthoryear{{Horne} et~al.,}{{Horne}
  et~al.}{2021}]{Horne2021}
{Horne} K.,  et~al., 2021, \mn@doi [\apj] {10.3847/1538-4357/abce60}, \href
  {https://ui.adsabs.harvard.edu/abs/2021ApJ...907...76H} {907, 76}

\bibitem[\protect\citeauthoryear{{Hunter}}{{Hunter}}{2007}]{Matplotlib}
{Hunter} J.~D.,  2007, \mn@doi [Computing in Science and Engineering]
  {10.1109/MCSE.2007.55}, \href
  {https://ui.adsabs.harvard.edu/abs/2007CSE.....9...90H} {9, 90}

\bibitem[\protect\citeauthoryear{{Kammoun}, {Papadakis}  \&
  {Dov{\v{c}}iak}}{{Kammoun} et~al.}{2019}]{Kammoun19lag}
{Kammoun} E.~S.,  {Papadakis} I.~E.,   {Dov{\v{c}}iak} M.,  2019, \mn@doi
  [\apjl] {10.3847/2041-8213/ab2a72}, \href
  {https://ui.adsabs.harvard.edu/abs/2019ApJ...879L..24K} {879, L24}

\bibitem[\protect\citeauthoryear{{Kammoun}, {Papadakis}  \&
  {Dov{\v{c}}iak}}{{Kammoun} et~al.}{2021a}]{Kammoun21b}
{Kammoun} E.~S.,  {Papadakis} I.~E.,   {Dov{\v{c}}iak} M.,  2021a, \mn@doi
  [\mnras] {10.1093/mnras/stab725}, \href
  {https://ui.adsabs.harvard.edu/abs/2021MNRAS.503.4163K} {503, 4163}

\bibitem[\protect\citeauthoryear{{Kammoun}, {Dov{\v{c}}iak}, {Papadakis},
  {Caballero-Garc{\'\i}a}  \& {Karas}}{{Kammoun} et~al.}{2021b}]{Kammoun21a}
{Kammoun} E.~S.,  {Dov{\v{c}}iak} M.,  {Papadakis} I.~E.,
  {Caballero-Garc{\'\i}a} M.~D.,   {Karas} V.,  2021b, \mn@doi [\apj]
  {10.3847/1538-4357/abcb93}, \href
  {https://ui.adsabs.harvard.edu/abs/2021ApJ...907...20K} {907, 20}

\bibitem[\protect\citeauthoryear{{Kara} et~al.,}{{Kara} et~al.}{2021}]{Kara21}
{Kara} E.,  et~al., 2021, \mn@doi [\apj] {10.3847/1538-4357/ac2159}, \href
  {https://ui.adsabs.harvard.edu/abs/2021ApJ...922..151K} {922, 151}

\bibitem[\protect\citeauthoryear{{Kara} et~al.,}{{Kara} et~al.}{2023}]{Kara23}
{Kara} E.,  et~al., 2023, \mn@doi [arXiv e-prints] {10.48550/arXiv.2302.07342},
  \href {https://ui.adsabs.harvard.edu/abs/2023arXiv230207342K} {p.
  arXiv:2302.07342}

\bibitem[\protect\citeauthoryear{{Lightman} \& {White}}{{Lightman} \&
  {White}}{1988}]{Lightman88}
{Lightman} A.~P.,  {White} T.~R.,  1988, \mn@doi [\apj] {10.1086/166905}, \href
  {https://ui.adsabs.harvard.edu/abs/1988ApJ...335...57L} {335, 57}

\bibitem[\protect\citeauthoryear{Matt, Perola  \& Piro}{Matt
  et~al.}{1991}]{Matt1991}
Matt G.,  Perola G.~C.,   Piro L.,  1991, Astronomy and Astrophysics, 247, 25

\bibitem[\protect\citeauthoryear{{McHardy} et~al.,}{{McHardy}
  et~al.}{2018}]{Mchardy18}
{McHardy} I.~M.,  et~al., 2018, \mn@doi [\mnras] {10.1093/mnras/sty1983}, \href
  {http://adsabs.harvard.edu/abs/2018MNRAS.480.2881M} {480, 2881}

\bibitem[\protect\citeauthoryear{{McHardy} et~al.,}{{McHardy}
  et~al.}{2023}]{McHardy23}
{McHardy} I.~M.,  et~al., 2023, \mn@doi [\mnras] {10.1093/mnras/stac3651},
  \href {https://ui.adsabs.harvard.edu/abs/2023MNRAS.519.3366M} {519, 3366}

\bibitem[\protect\citeauthoryear{{Netzer}}{{Netzer}}{2022}]{Netzer2022}
{Netzer} H.,  2022, \mn@doi [\mnras] {10.1093/mnras/stab3133}, \href
  {https://ui.adsabs.harvard.edu/abs/2022MNRAS.509.2637N} {509, 2637}

\bibitem[\protect\citeauthoryear{{Novikov} \& {Thorne}}{{Novikov} \&
  {Thorne}}{1973}]{Novikov73}
{Novikov} I.~D.,  {Thorne} K.~S.,  1973, in {Dewitt} C.,  {Dewitt} B.~S.,  eds,
  Black Holes (Les Astres Occlus). pp 343--450

\bibitem[\protect\citeauthoryear{{Pahari}, {McHardy}, {Vincentelli}, {Cackett},
  {Peterson}, {Goad}, {G{\"u}ltekin}  \& {Horne}}{{Pahari}
  et~al.}{2020}]{Pahari20}
{Pahari} M.,  {McHardy} I.~M.,  {Vincentelli} F.,  {Cackett} E.,  {Peterson}
  B.~M.,  {Goad} M.,  {G{\"u}ltekin} K.,   {Horne} K.,  2020, \mn@doi [\mnras]
  {10.1093/mnras/staa1055}, \href
  {https://ui.adsabs.harvard.edu/abs/2020MNRAS.494.4057P} {494, 4057}

\bibitem[\protect\citeauthoryear{{Panagiotou}, {Papadakis}, {Kammoun}  \&
  {Dov{\v{c}}iak}}{{Panagiotou} et~al.}{2020}]{Panagiotou20}
{Panagiotou} C.,  {Papadakis} I.~E.,  {Kammoun} E.~S.,   {Dov{\v{c}}iak} M.,
  2020, \mn@doi [\mnras] {10.1093/mnras/staa2920}, \href
  {https://ui.adsabs.harvard.edu/abs/2020MNRAS.499.1998P} {499, 1998}

\bibitem[\protect\citeauthoryear{{Panagiotou}, {Papadakis}, {Kara}, {Kammoun}
  \& {Dov{\v{c}}iak}}{{Panagiotou} et~al.}{2022}]{Panagiotou22}
{Panagiotou} C.,  {Papadakis} I.,  {Kara} E.,  {Kammoun} E.,   {Dov{\v{c}}iak}
  M.,  2022, \mn@doi [\apj] {10.3847/1538-4357/ac7e4d}, \href
  {https://ui.adsabs.harvard.edu/abs/2022ApJ...935...93P} {935, 93}

\bibitem[\protect\citeauthoryear{{Pozo Nu{\~n}ez} et~al.,}{{Pozo Nu{\~n}ez}
  et~al.}{2019}]{Pozonunez2019}
{Pozo Nu{\~n}ez} F.,  et~al., 2019, \mn@doi [\mnras] {10.1093/mnras/stz2830},
  \href {https://ui.adsabs.harvard.edu/abs/2019MNRAS.490.3936P} {490, 3936}

\bibitem[\protect\citeauthoryear{{Ross}, {Fabian}  \& {Mineshige}}{{Ross}
  et~al.}{1992}]{Ross92}
{Ross} R.~R.,  {Fabian} A.~C.,   {Mineshige} S.,  1992, \mn@doi [\mnras]
  {10.1093/mnras/258.1.189}, \href
  {https://ui.adsabs.harvard.edu/abs/1992MNRAS.258..189R} {258, 189}

\bibitem[\protect\citeauthoryear{{Shakura} \& {Sunyaev}}{{Shakura} \&
  {Sunyaev}}{1973}]{Shakura73}
{Shakura} N.~I.,  {Sunyaev} R.~A.,  1973, \aap, \href
  {http://adsabs.harvard.edu/abs/1973A%26A....24..337S} {24, 337}

\bibitem[\protect\citeauthoryear{{Shimura} \& {Takahara}}{{Shimura} \&
  {Takahara}}{1995}]{Shimura95}
{Shimura} T.,  {Takahara} F.,  1995, \mn@doi [\apj] {10.1086/175740}, \href
  {https://ui.adsabs.harvard.edu/abs/1995ApJ...445..780S} {445, 780}

\bibitem[\protect\citeauthoryear{{Vasudevan} \& {Fabian}}{{Vasudevan} \&
  {Fabian}}{2007}]{Vasudevan07}
{Vasudevan} R.~V.,  {Fabian} A.~C.,  2007, \mn@doi [\mnras]
  {10.1111/j.1365-2966.2007.12328.x}, \href
  {https://ui.adsabs.harvard.edu/abs/2007MNRAS.381.1235V} {381, 1235}

\bibitem[\protect\citeauthoryear{{Vincentelli} et~al.,}{{Vincentelli}
  et~al.}{2021}]{Vincentelli21}
{Vincentelli} F.~M.,  et~al., 2021, \mn@doi [\mnras] {10.1093/mnras/stab1033},
  \href {https://ui.adsabs.harvard.edu/abs/2021MNRAS.504.4337V} {504, 4337}

\bibitem[\protect\citeauthoryear{{Vincentelli}, {McHardy}, {Hern{\'a}ndez
  Santisteban}, {Cackett}, {Gelbord}, {Horne}, {Miller}  \&
  {Lobban}}{{Vincentelli} et~al.}{2022}]{Vincentelli22}
{Vincentelli} F.~M.,  {McHardy} I.,  {Hern{\'a}ndez Santisteban} J.~V.,
  {Cackett} E.~M.,  {Gelbord} J.,  {Horne} K.,  {Miller} J.~A.,   {Lobban} A.,
  2022, \mn@doi [\mnras] {10.1093/mnrasl/slac009}, \href
  {https://ui.adsabs.harvard.edu/abs/2022MNRAS.512L..33V} {512, L33}

\bibitem[\protect\citeauthoryear{{Virtanen} et~al.,}{{Virtanen}
  et~al.}{2020}]{Scipy}
{Virtanen} P.,  et~al., 2020, \mn@doi [Nature Methods]
  {10.1038/s41592-019-0686-2}, \href
  {https://ui.adsabs.harvard.edu/abs/2020NatMe..17..261V} {17, 261}

\bibitem[\protect\citeauthoryear{{Wales} \& {Doye}}{{Wales} \&
  {Doye}}{1997}]{Wales97}
{Wales} D.~J.,  {Doye} J. P.~K.,  1997, \mn@doi [Journal of Physical Chemistry
  A] {10.1021/jp970984n}, \href
  {https://ui.adsabs.harvard.edu/abs/1997JPCA..101.5111W} {101, 5111}

\makeatother
\end{thebibliography}


\appendix
\section{Time lag spectra fits}
\label{sec:appendix_figures}
\begin{figure}
    {\centering
    \includegraphics[width=0.85\linewidth]{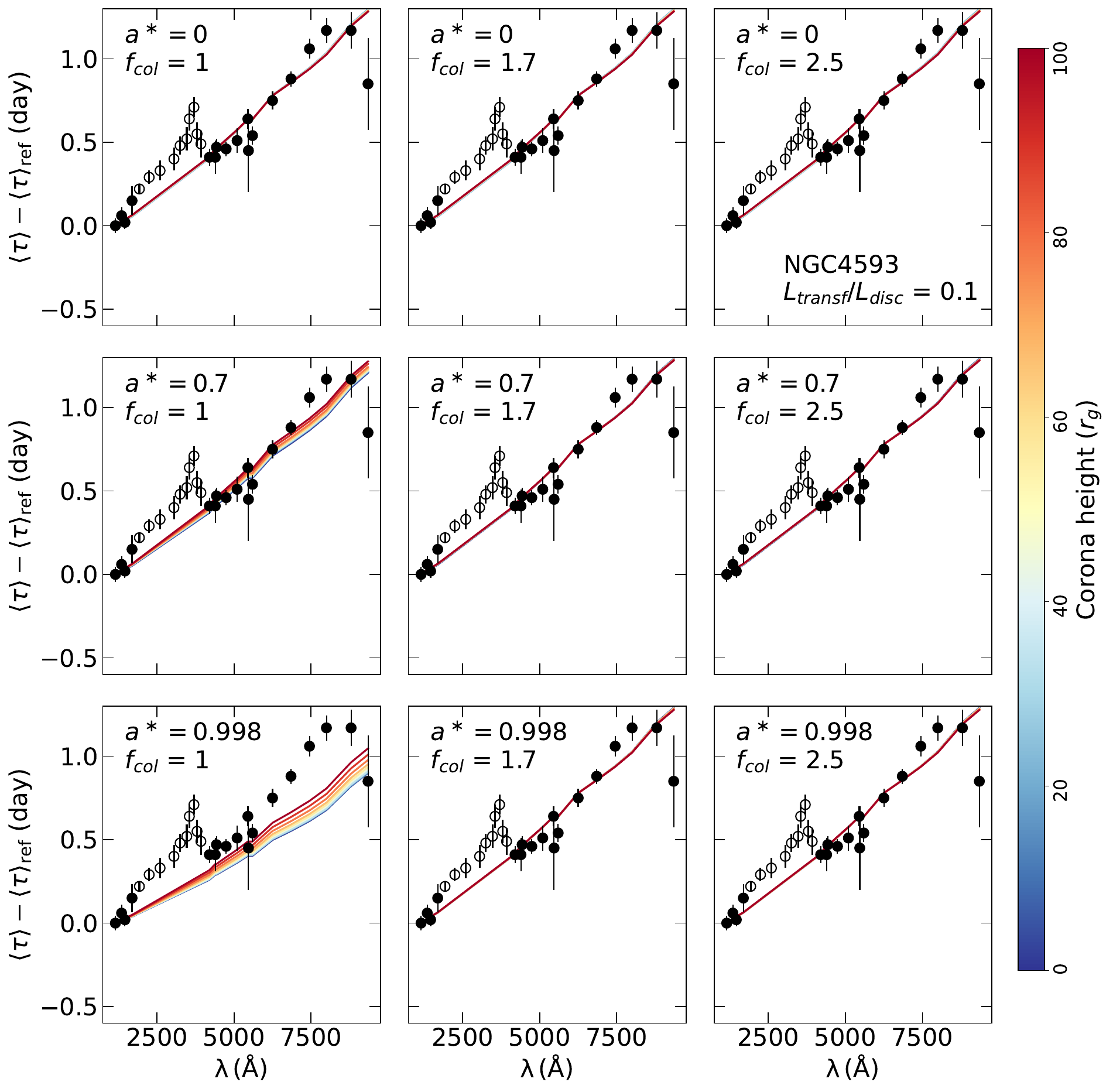}
    \includegraphics[width=0.85\linewidth]{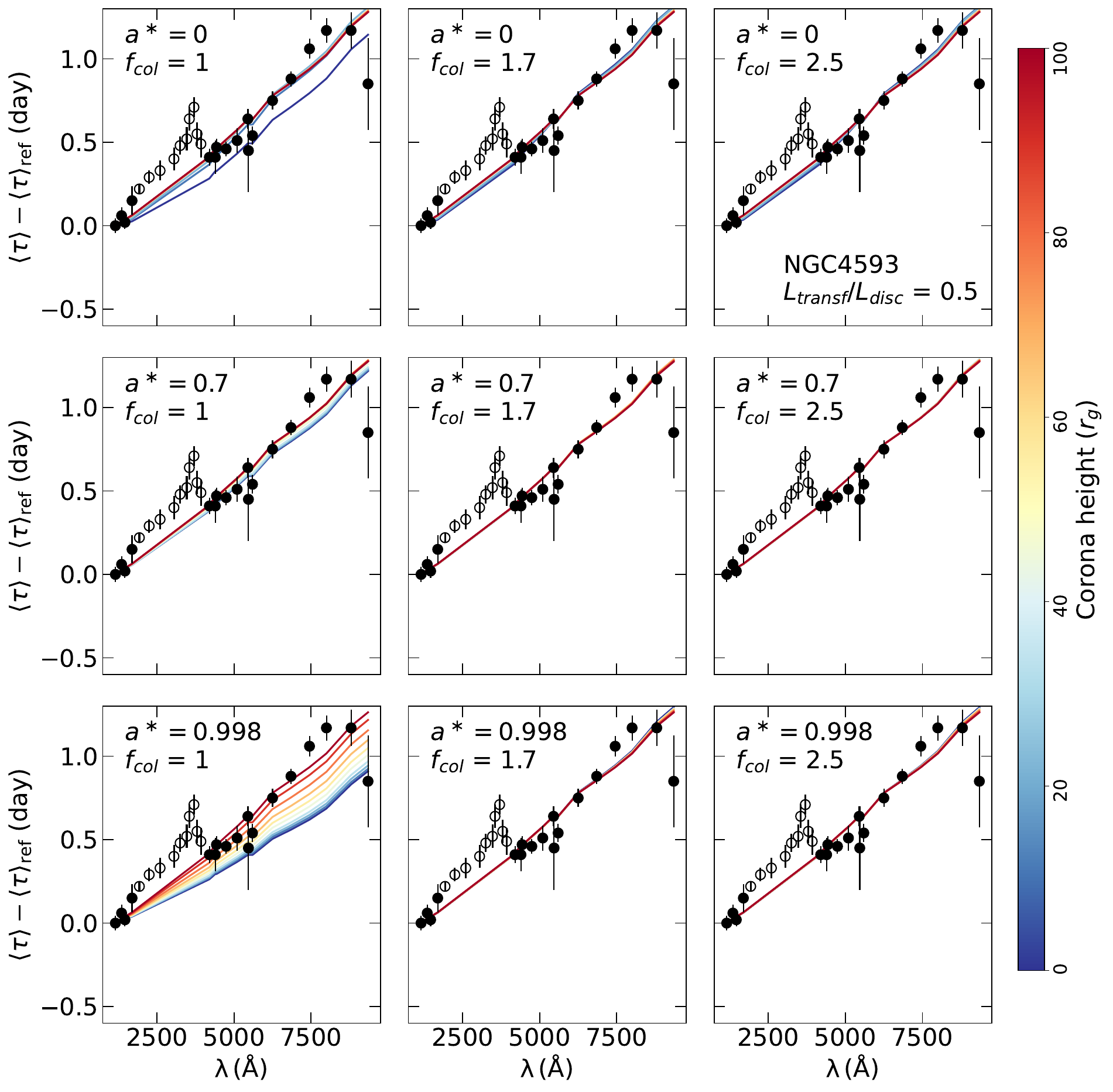}
    \includegraphics[width=0.85\linewidth]{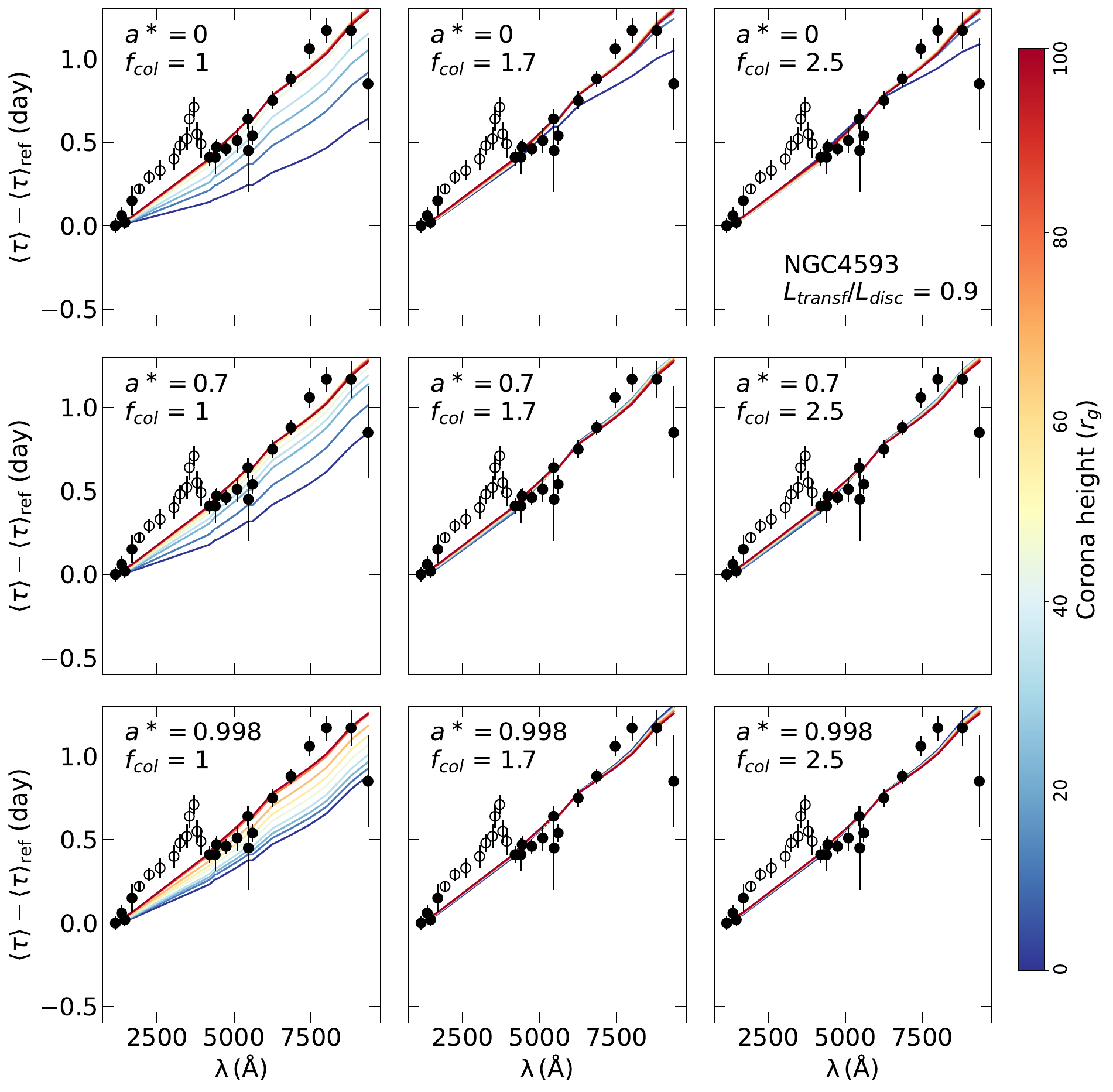}
    }
\caption{Fitting the time-lag spectra of NGC\,4593 for different values of \ltransf. We assume $a^\ast = 0, 0.7,$ and 0.998 (top, middle, and bottom rows, respectively) and $\fcol = 1, 1.7,$ and 2.5 (left to right columns). The color maps correspond to the different values of the coronal height. Open circles show time-lag measurements that we did not consider when fitting the data, as they may be affected by emission from the much larger (than the disc) BLR.}
\label{fig:timelag_NGC4593}
\end{figure}

\begin{figure}
    {\centering
    \includegraphics[width=0.85\linewidth]{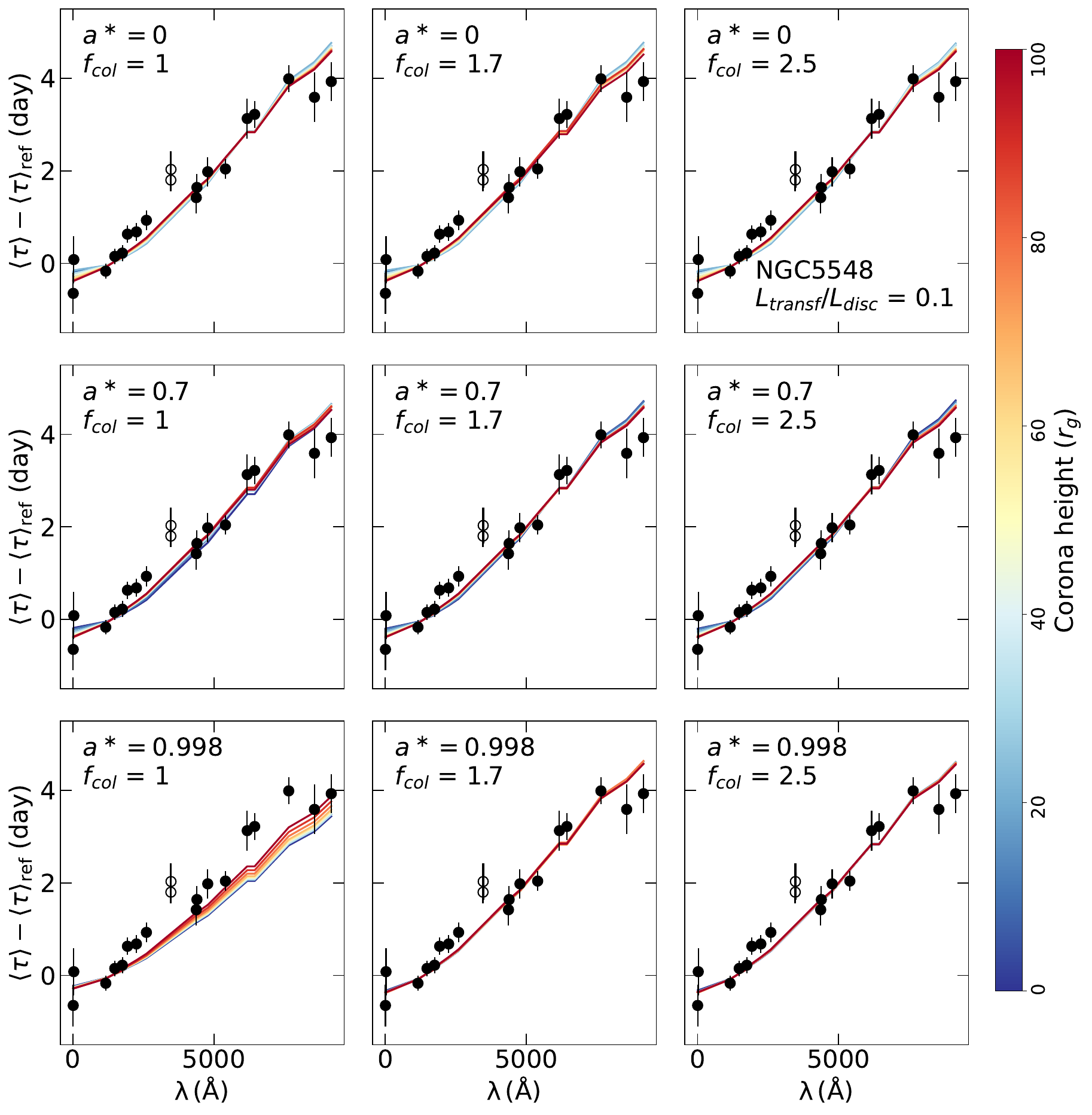}
    \includegraphics[width=0.85\linewidth]{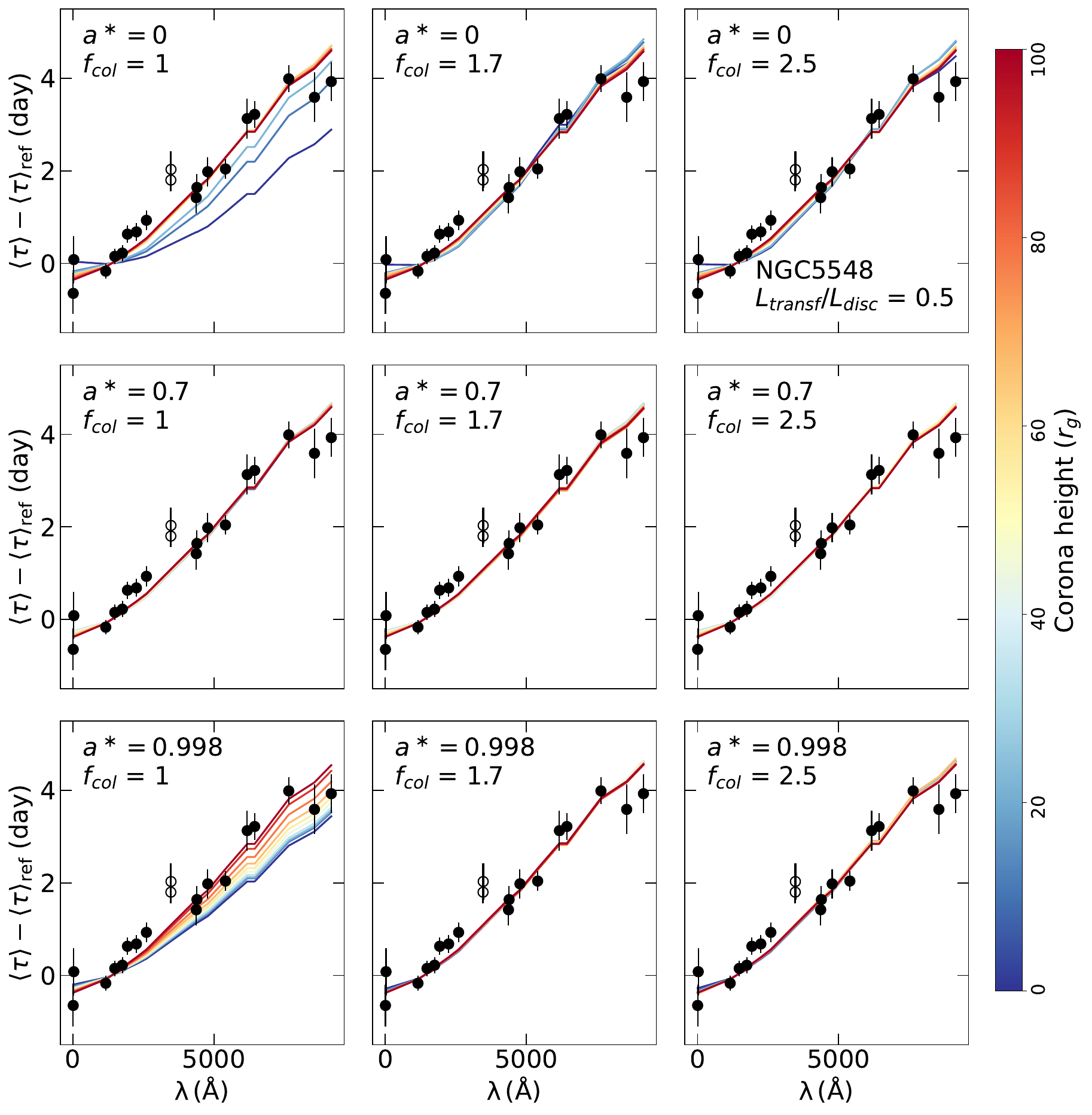}
    \includegraphics[width=0.85\linewidth]{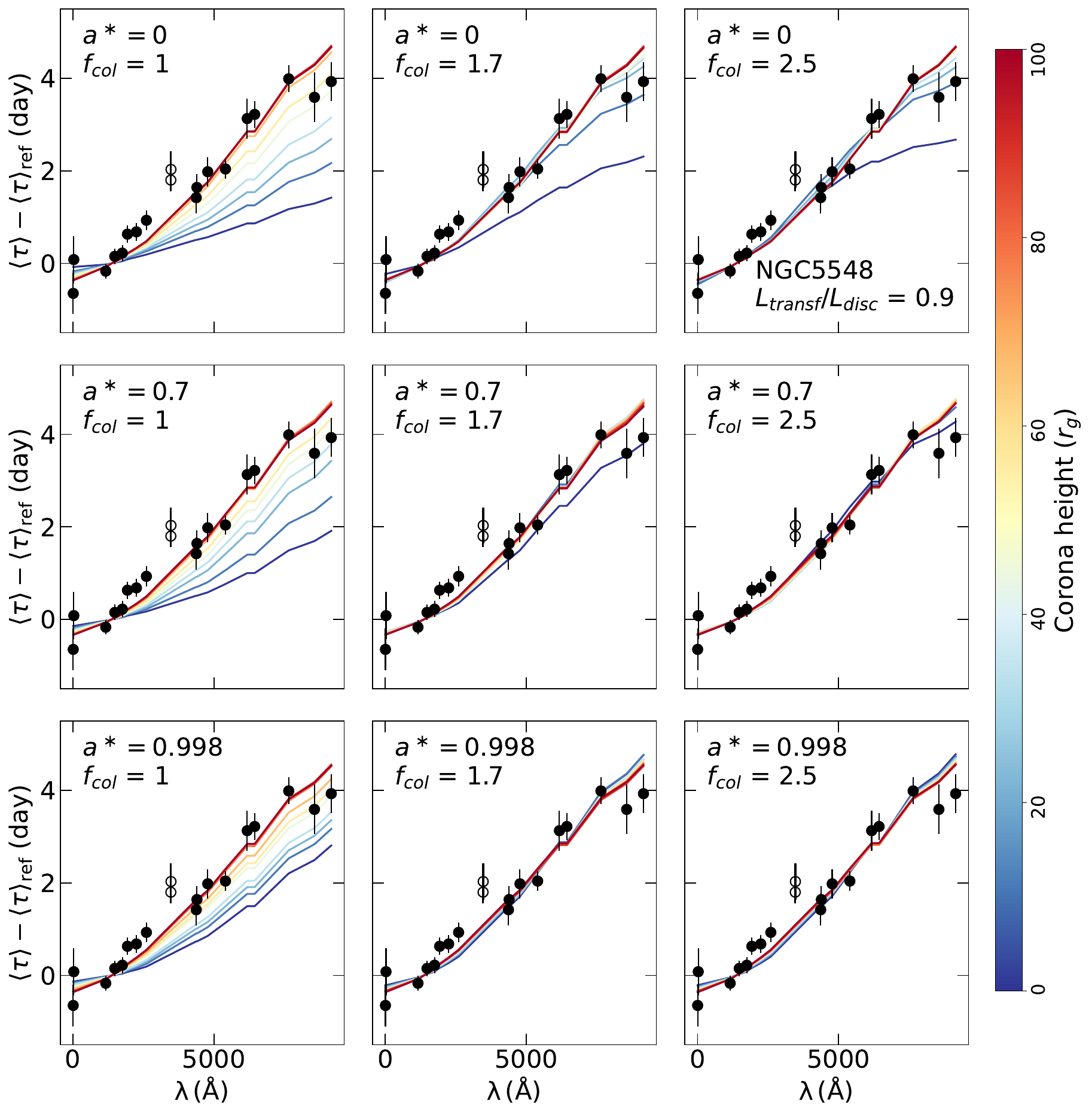}
    }
\caption{Same as Fig.\,\ref{fig:timelag_NGC4593} but for NGC\,5548.}
\label{fig:timelag_NGC5548}
\end{figure}

\begin{figure}
    {\centering
    \includegraphics[width=0.85\linewidth]{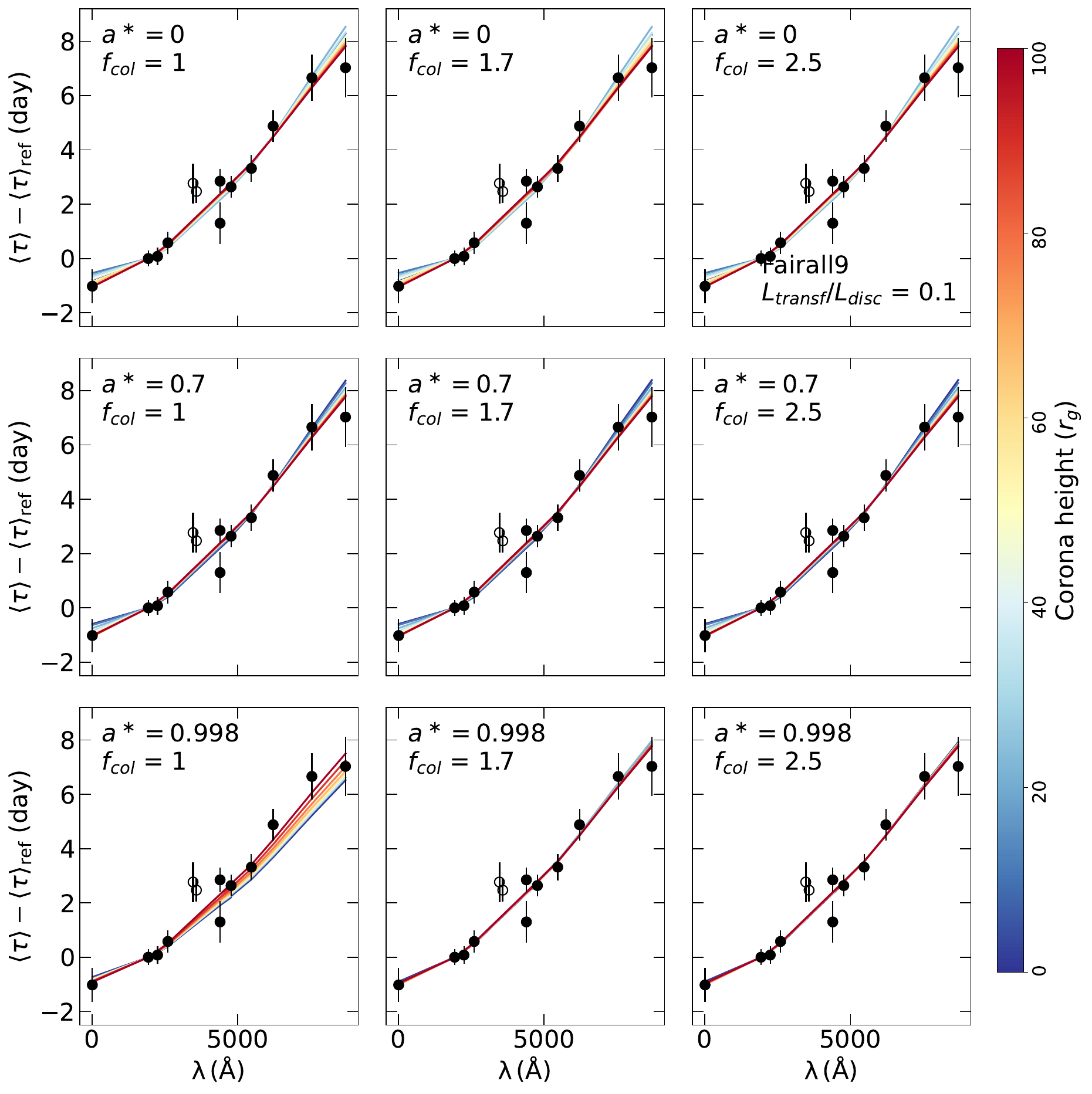}
    \includegraphics[width=0.85\linewidth]{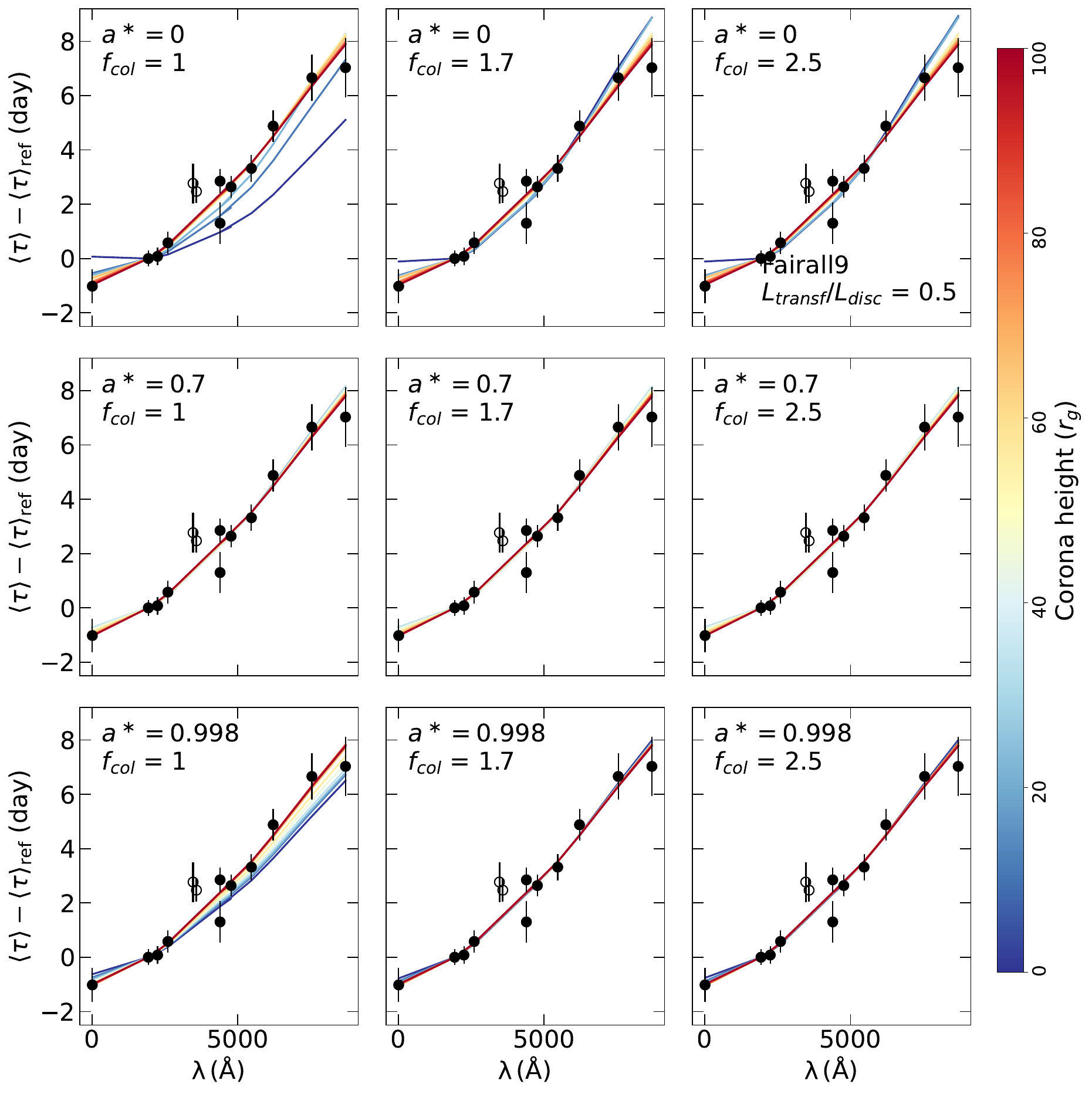}
    \includegraphics[width=0.85\linewidth]{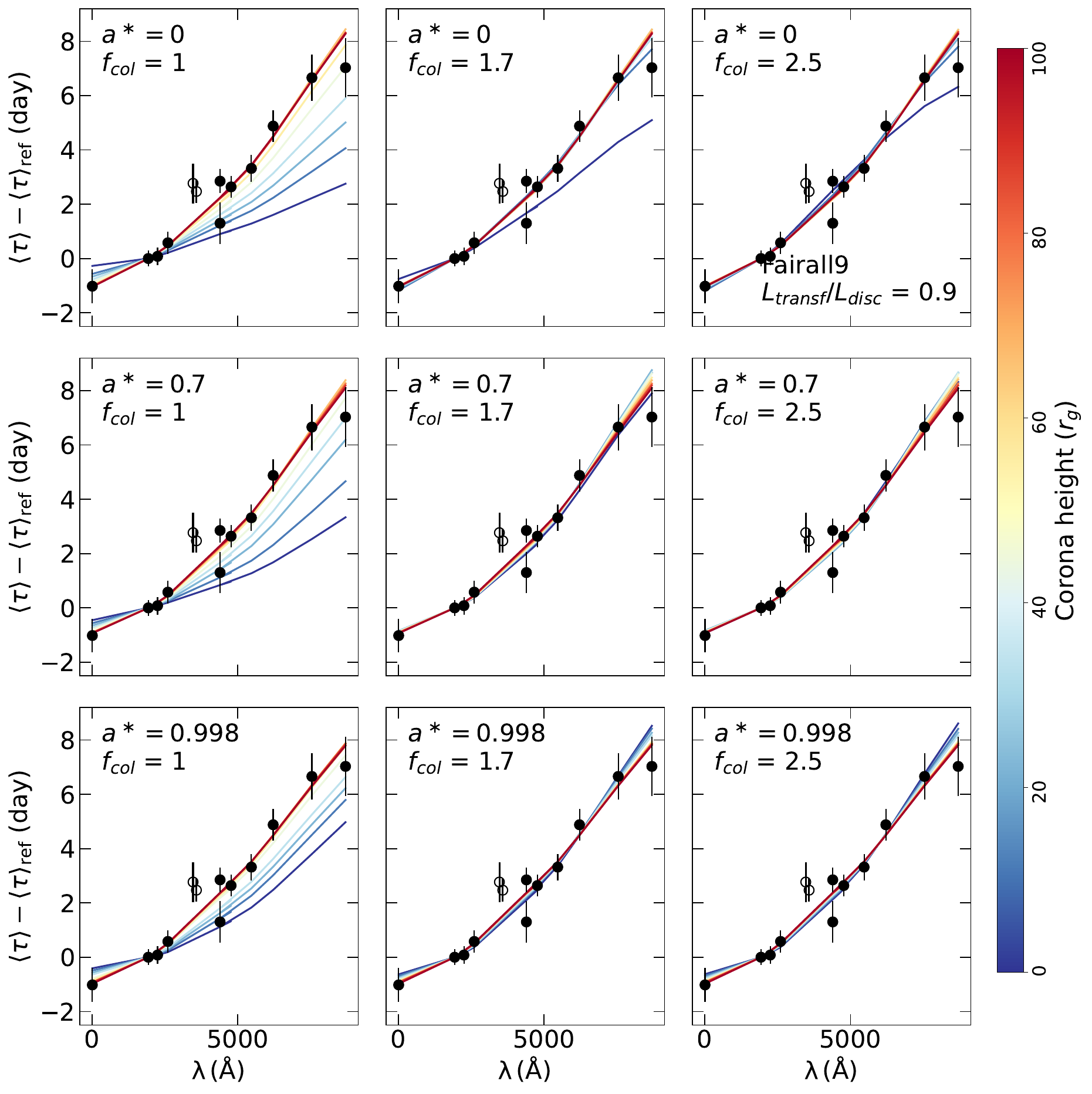}
    }
\caption{Same as Fig.\,\ref{fig:timelag_NGC4593} but for Fairall\,9.}
\label{fig:timelag_F9}
\end{figure}

\begin{figure}
    {\centering
    \includegraphics[width=0.85\linewidth]{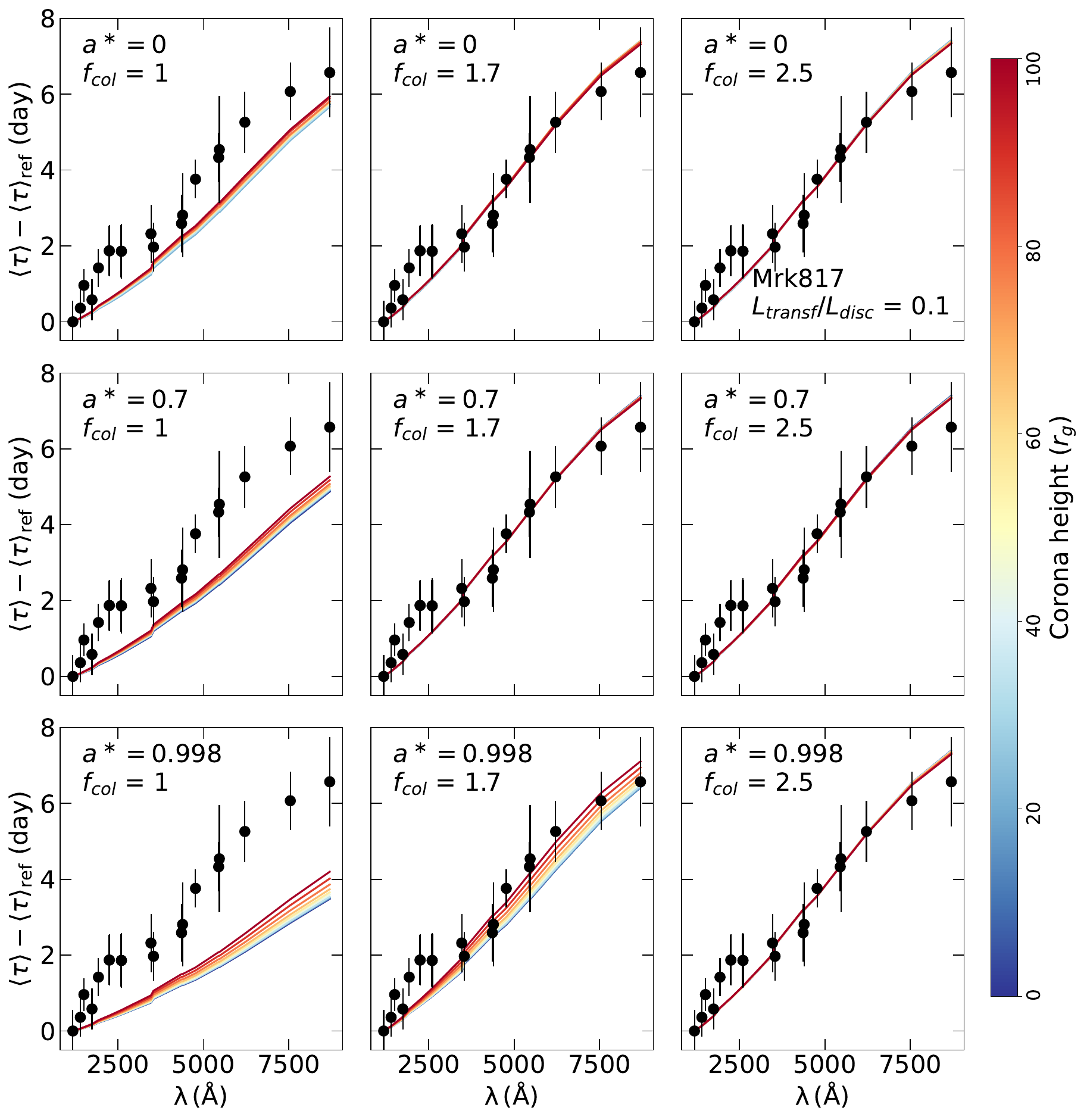}
    \includegraphics[width=0.85\linewidth]{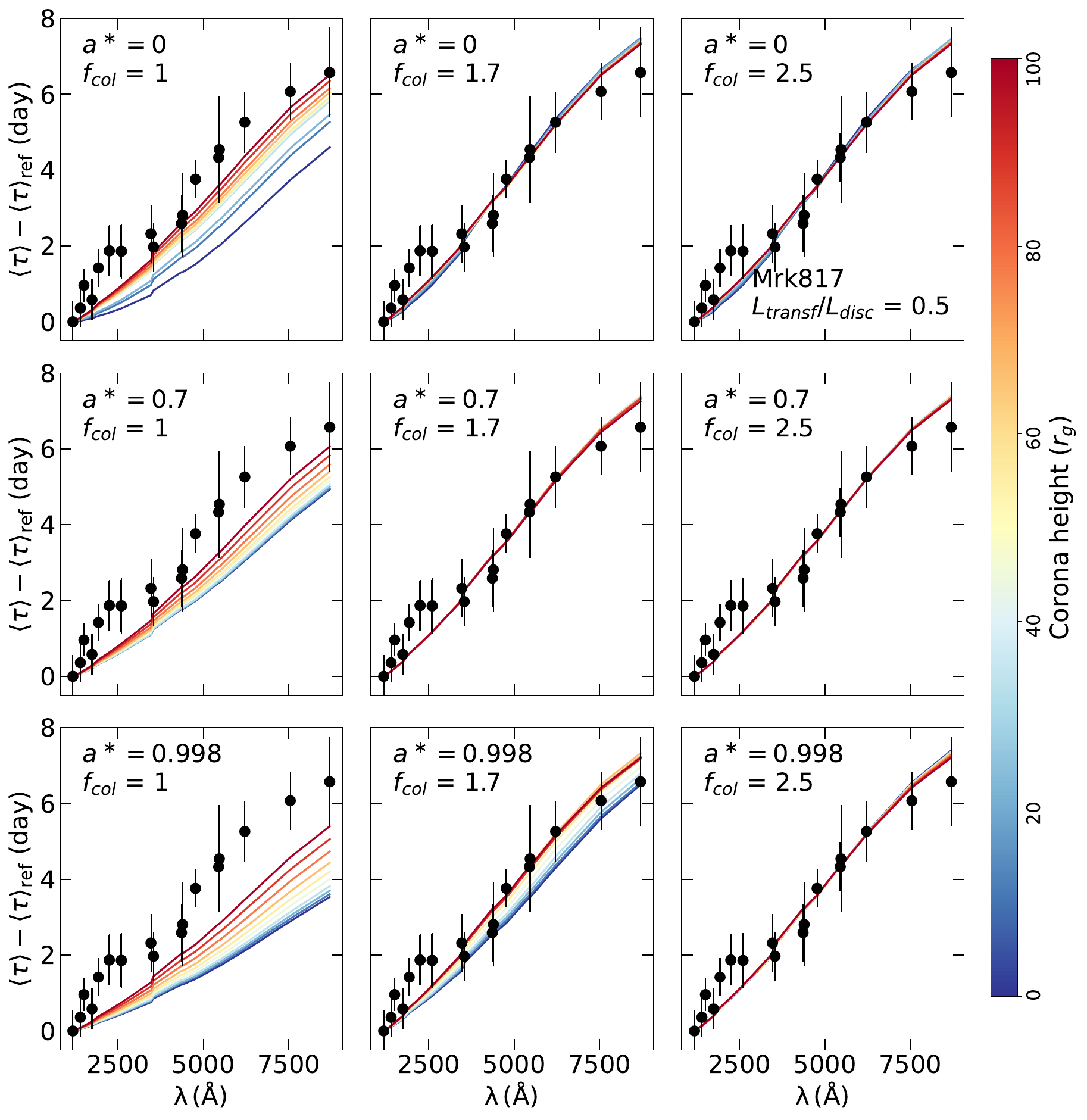}
    \includegraphics[width=0.85\linewidth]{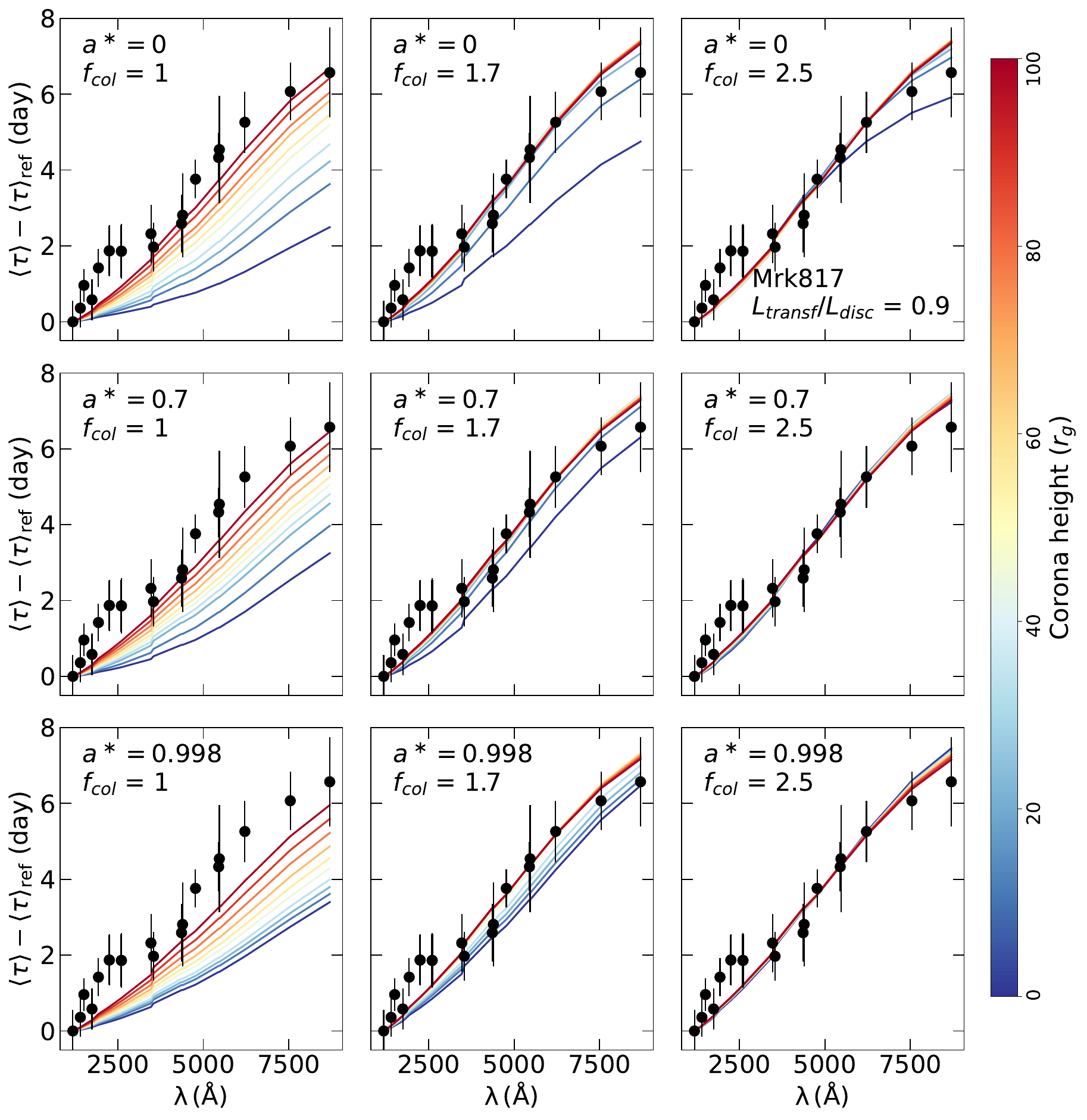}
    
\caption{Same as Fig.\,\ref{fig:timelag_NGC4593} but for Mrk\,817.}
}
\label{fig:timelag_Mrk817}
\end{figure}

\newpage

\section{Updating the analytic prescription}
\label{sec:analytic}
K21b derived an analytic expression of the time lag spectra as a function of the BH mass, corona height, observed $2-10\,\rm keV$ luminosity, and the accretion rate, for spins of 0 and 0.998. The resulting functions were computed by assuming that the X--ray source is powered via an external process that is not linked to accretion (equivalent to a negative value of \ltransf\ in the current code) and for $\fcol = 2.4$. In this section, we 
provide a new term that can be included in the K21b equations to take into account the use of \fcol\ values different than 2.4. In addition, we compare the time-lags when we assume an in internally and an externally powered X--ray corona, and we comment on how the analytical functions could be scaled to provide time-lag estimates in the former case. 

\subsection{The dependence of time-lags on \fcol}

\begin{figure*}
    \centering
    \includegraphics[width=0.99\linewidth]{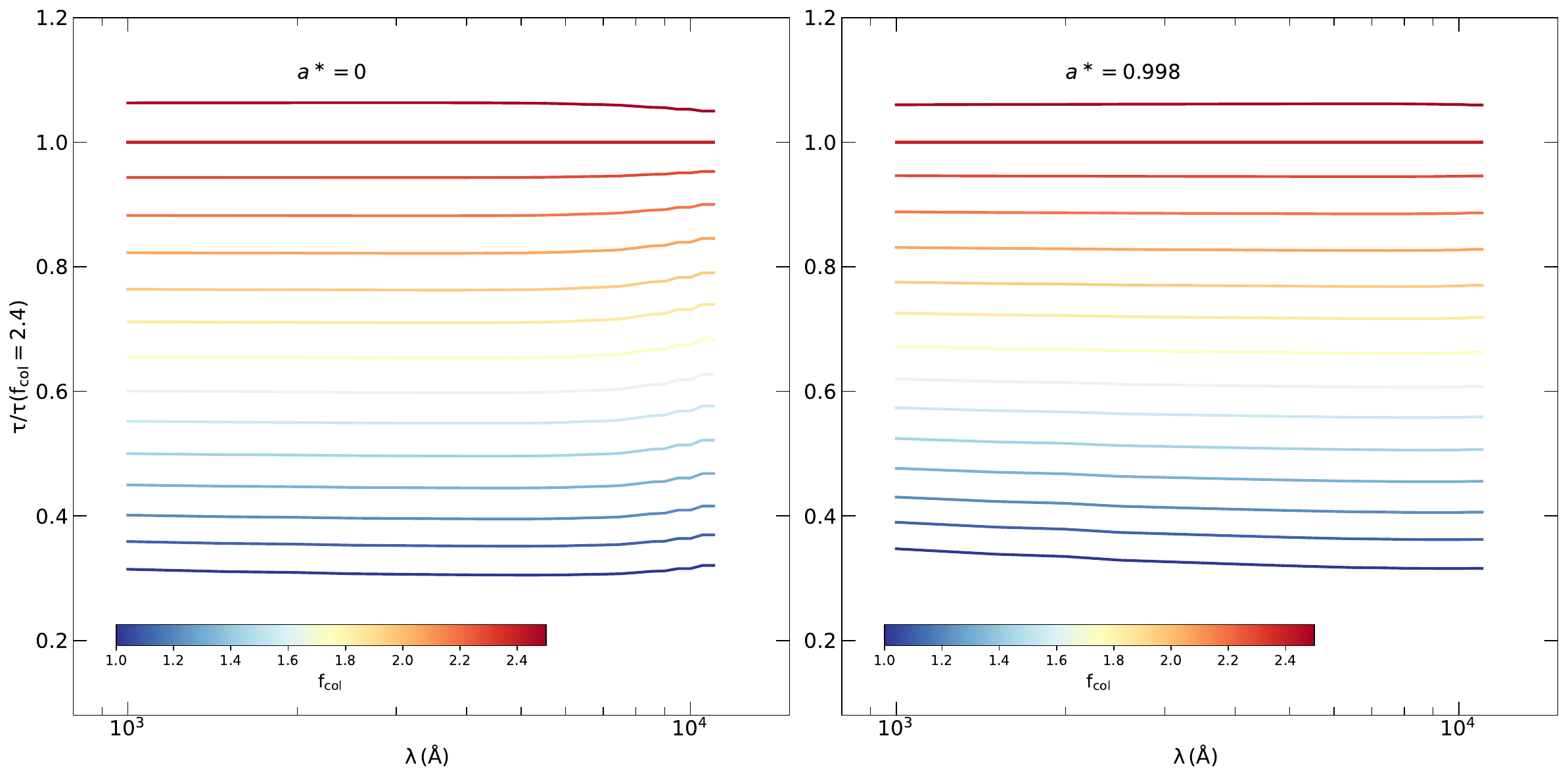}
    \caption{Ratio of the time lags over the time lag assuming $\fcol = 2.4$ as a function of wavelength for $a^\ast = 0$ and 0.998 (left and right panels, respectively). The color code corresponds to the values of $\fcol$.}
    \label{fig:ratio_vslambda_fcol}
\end{figure*}

\begin{figure}
    \centering
    \includegraphics[width=0.99\linewidth]{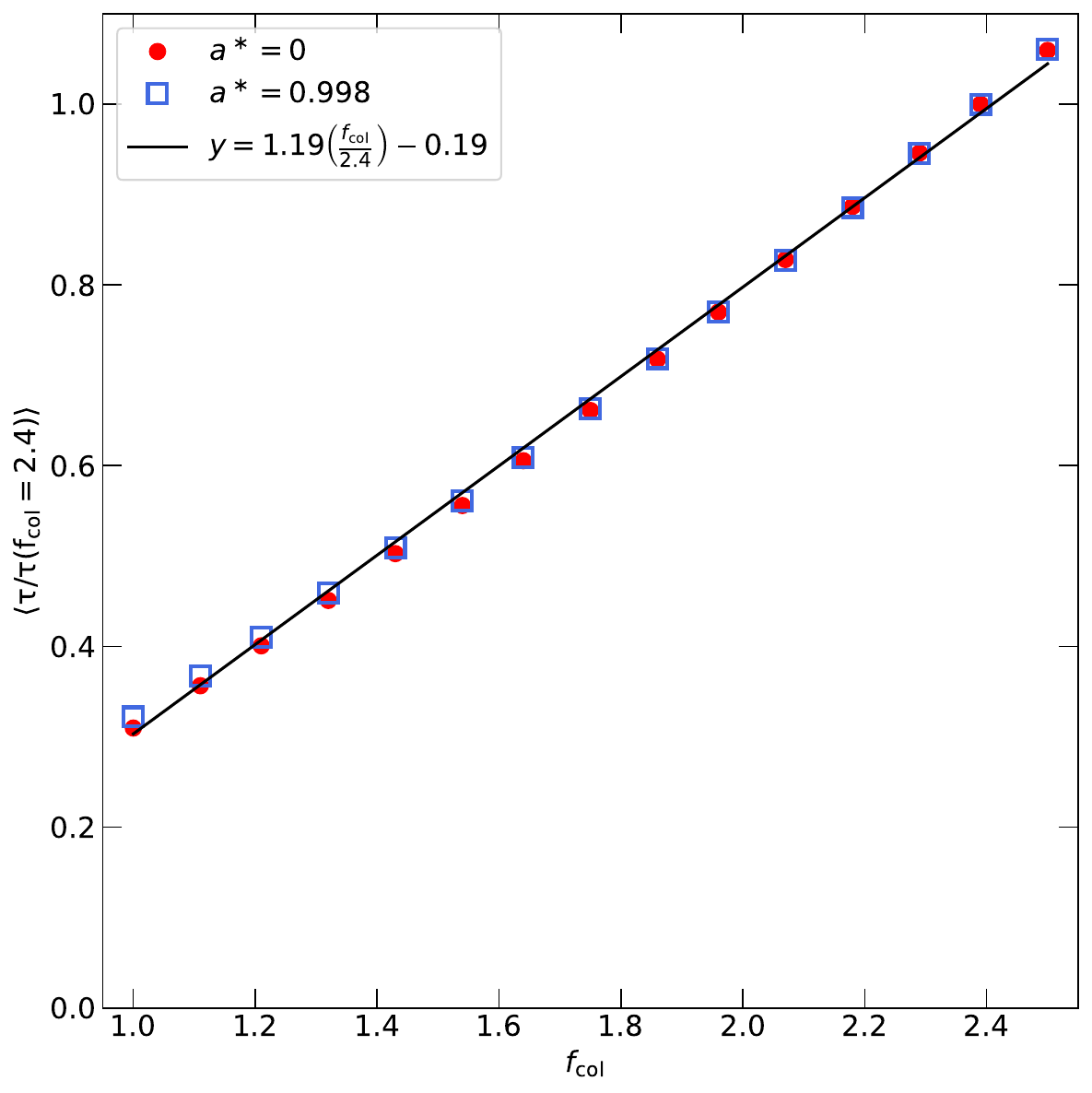}
    \caption{Average ratio of the time lags over the time lag assuming $\fcol = 2.4$ as a function of \fcol, $a^\ast = 0$ and 0.998 (filled circles and empty squares, respectively). The solid black line correspond to the best-fitting linear relation to the data points.}
    \label{fig:ratio_vs_fcol}
\end{figure}

In this section, we investigate the dependency of the time lags on \fcol, in the case of externally powered corona. We performed a set of simulations assuming the same fiducial parameters as K21b, but for a range of \fcol\ between 1 and 2.5, for $a^\ast =0$ and 0.998. We assumed, in all of them, $\ltransf = -0.5$. We remind the reader that the analytic expression presented by K21b was estimated using $\fcol = 2.4$. Figure\,\ref{fig:ratio_vslambda_fcol} shows the ratio of the time lag for a given value of \fcol\ divided by the time lag for $\fcol = 2.4$ as a function of wavelength for $a^\ast = 0$ and 0.998 (left and right panels, respectively). This  figure shows that the time lags increase as \fcol\ increases, by the same factor at all wavelengths (i.e., the ratios are constant as function of wavelength). In addition, these ratios are constant with spin. 
Figure\,\ref{fig:ratio_vs_fcol} shows the ratio of the time-lags for \fcol$\neq 2.4)$ over the time-lags for \fcol=2.4. The solid line shows that a straight line gives a very good fit to the data (both for spin 0 and 0.998). This result allows us to update the analytic expression derived by the K21b as follows:

\begin{eqnarray}
\tau (\lambda) - \tau (\lambda_{\rm R}) &=& A(h_{10}) M_7^{0.7} f_1(\dot{m}_{0.05})f_2(L_{\rm X,0.01}) f_3(f_{\rm col}) \nonumber \\
& \times &\left(\lambda_{1950}^{B(h_{10})} - \lambda_{\rm R,1950}^{B(h_{10})} \right)~{\rm day}, 
\label{timelagseq}
\end{eqnarray}

\noindent where $\tau(\lambda)$ and $\tau(\lambda_{\rm R})$ are the time-lags between X--rays and the bands at $\lambda$ and $\lambda_{\rm R}$, respectively ($\lambda_{1950}=\lambda/1950$~\AA, and $\lambda_{\rm R, 1950}=\lambda_{\rm R}/1950$\AA). In the equation above, $h_{10}$ is the height of the lamp-post in units of 10~\rg, $M_7$ is the BH mass in units of $10^7 M_\odot$, $\dot{m}_{0.05}$ is the accretion rate in units of 5$\%$ of the Eddington luminosity, $L_{\rm X,0.01}$ is the observed, $2-10$~keV luminosity in units of 0.01 of the Eddington luminosity. Functions $A(h_{10}), B(h_{10}), f_1(\dot{m}_{0.05}),$ and $f_2(L_{X,0.01})$, are given by Eqs.\,9-12 and 13-16, for spin 0 and 1, in K21b. As for the \fcol, the term, 

\begin{equation}
    f_3(\fcol) = 1.19 \left( \frac{\fcol}{2.4}\right) - 0.19,
\end{equation}

\noindent is the same for both $a^\ast = 0$ and 0.998. This is the equation for the best-fit line plotted in Fig.\,\ref{fig:ratio_vs_fcol}.

\subsection{Effects of \ltransf}

We repeated the simulations as K21b by choosing the same fiducial parameters of $M_{\rm BH}, \mdot, h, \theta, \Gamma,$ and $E_{\rm cut}$. We also considered two values of the spin, 0 and 0.998, and two values of \fcol\ being 1 and 2.4. 

\begin{figure}
    \centering
    \includegraphics[width=0.99\linewidth]{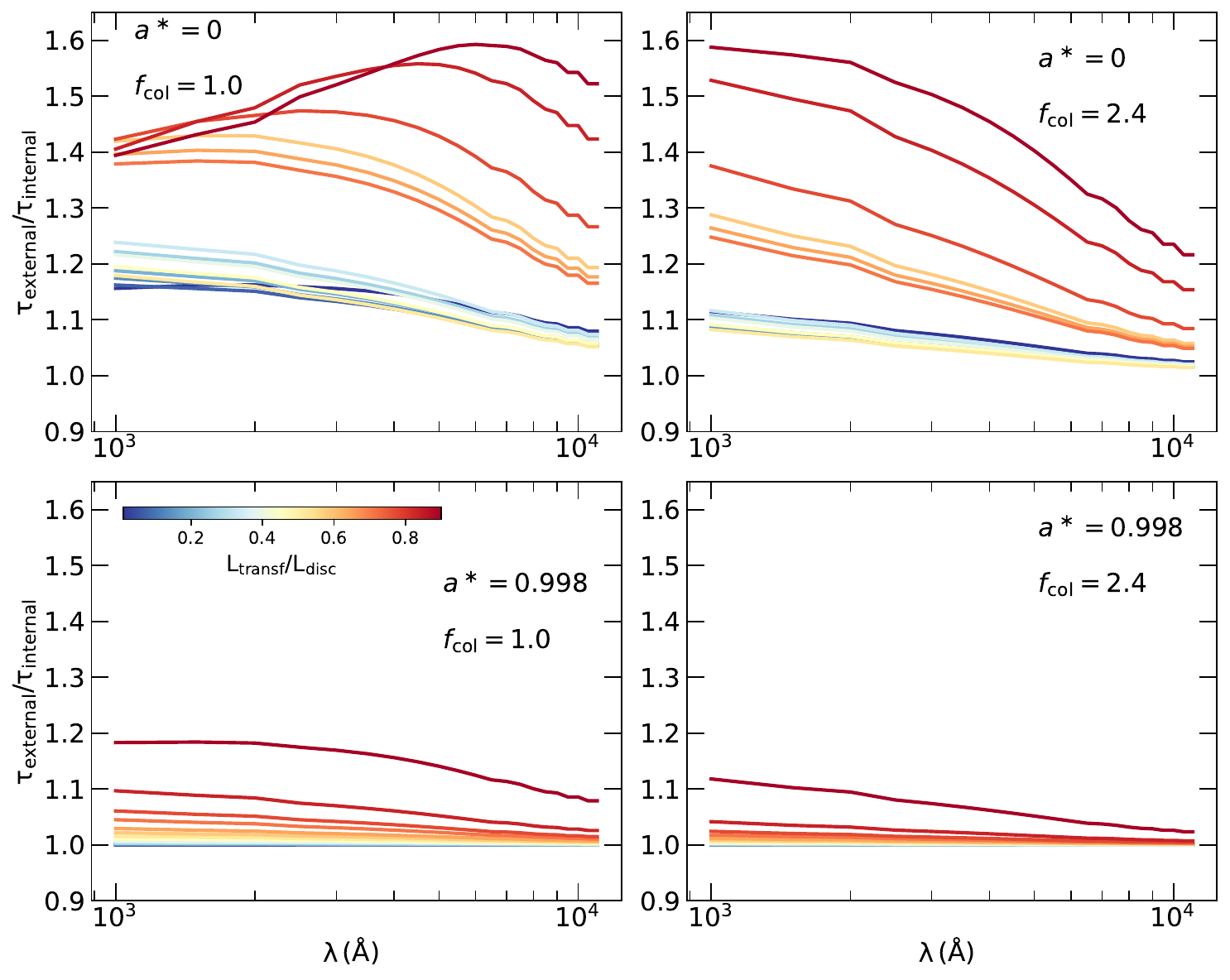}
    \caption{Ratio of the time lag assuming external power of the X--ray source over time lag assuming an X--ray source powered by the accretion flow as a function of wavelength for $a^\ast = 0$ and 0.998 (top and bottom panels, respectively) and $\fcol = 1$ and 2.4 (left and right panels, respectively). The color code correspond to the values of $|\ltransf|$. }
    \label{fig:ratio_Ltransf}
\end{figure}

Figure\,\ref{fig:ratio_Ltransf} shows the ratios of the time lags for external power corona divided by the time lags for internal power,  for the different combinations of spin and \fcol. In all cases, the time lags of externally powered corona are larger than the ones of internally powered corona, as already mentioned in Section\,\ref{sec:Ltransfdep}.  The difference in shape and amplitude depends on the chosen values of spin and \fcol. In the case of $a^\ast = 0.998$, the time-lags difference is smaller than 20\%, for any value of \fcol. Most probably, such differences cannot be detected with the currently available time-lag spectra. We conclude that Eq.\,(\ref{timelagseq}) above works well, both in the case of externally and the case of the internally powered corona, in the case of high spins. 

The upper panels in Fig.\ref{fig:ratio_Ltransf} show that for \ltransf\ values smaller than $\sim 0.4$, the time-lags difference is smaller than $\sim 20$\%. We therefore conclude that, even in the case of zero spin,  Eq.\,(\ref{timelagseq}) should provide the correct time-lags, irrespective of how the X--ray corona is powered, as long as \ltransf\ is smaller than $\sim 0.4$. However, this is not the case when \ltransf\ increases. In this case, the shape of the time-lag spectra changes, and the differences depend on \fcol\, as well. In general, we can see that the time lags can change by a factor of up to $\sim 1.6$ at long wavelengths, when $\fcol=1$, and by the same factor at small wavelengths, when $\fcol=2.4$ (and $\ltransf=0.9$).  

We cannot find a straightforward prescription to add into Eq.\,(\ref{timelagseq}) that could take into account the assumption of an externally or internally powered corona that will be valid for $\ltransf$ values larger than $\sim 0.4$. For that reason, we recommend the use of the code we present in this paper to model the observed time-lags, under the desired assumption regarding the source of power for the X--ray corona.

\bsp	
\label{lastpage}
\end{document}